\documentclass[%
 reprint,
 amsmath,amssymb,
 aps,
prd]{revtex4-2}

\usepackage{booktabs}
\usepackage{graphicx}
\usepackage{dcolumn}
\usepackage{bm}
\usepackage{lineno}
\usepackage{braket}
\usepackage{multirow}
\usepackage{hyperref}

\usepackage{xspace} 

\def\beq{\begin{equation}}
\def\eeq{\end{equation}}

\def\CP{$CP$ }
\def\C      {\ensuremath{C}\xspace}

\def\psipp{\psi(3770)}

\def\DSTO{\ensuremath{D^{*0}}\xspace}

\def\aDSTO{\bar{D}^{*0}}

\def\DSTpDSTm{\ensuremath{D^{*+}}\ensuremath{D^{*-}}\xspace}

\def\DD{\ensuremath{\D\Db\;}\xspace}
\def\DSTDG{\ensuremath{D^{*}\bar{D}\to {\gamma} \D\Db}\xspace}
\def\DSTDP{\ensuremath{D^{*}\bar{D}\to{\pi^0} \D \Db}\xspace}
\def\DSTDSTEven{\ensuremath{D^{*}\bar{D}^{*}\to{\gamma\pi^0}\D \Db}\xspace}
\def\DSTDSTOdd{\ensuremath{D^{*}\bar{D}^{*}\to{\gamma\gamma/\pi^0\pi^0}\D \Db}\xspace}

\def\DSTD{\ensuremath{\Dstar\Db\xspace}}

\def\DSTDST{\ensuremath{\Dstar\Dstarb}\xspace}

\def\ee{\ensuremath{e^+e^-}\xspace}

\def\Ecm{E_{\text{cm}}}

\def\Emiss{E_{\text{miss}}}
\def\MMSq{M_{\text{miss,$\DD$}}^2}
\def\MMSqGamma{M_{\text{miss,$\gamma\DD$}}^2}
\def\DStarMRec{M_{\text{rec,}\Dstar}}
\newcommand{\MRec}[1]{M_{\text{rec,}#1}}
\def\DeltaMDST{\Delta M_{\Dstar}}

\def\DeltaMrec{\Delta\MRec{\D}}
\def\pslow{p_{\pi,\text{slow}}}

\def\invfb{\text{ fb}^{-1}}
\def\MeV{\text{ MeV}}
\def\GeV{\text{ GeV}}
\def\MeVcc{\text{ MeV}/c^2}

\def\kaon    {{\ensuremath{\PK}}\xspace}
\def\KS      {{\ensuremath{K^0_{\mathrm{S}}}}\xspace}
\def\K      {{\ensuremath{\kaon}}\xspace}

\def\Ki      {{\ensuremath{K_{i}}}\xspace}
\def\Kmi      {{\ensuremath{K_{-i}}}\xspace}
\def\ci      {{\ensuremath{c_{i}}}\xspace}
\def\si      {{\ensuremath{s_{i}}}\xspace}

\def\thebaroffset{0.1em}
\newcommand{\offsetoverline}[2][\thebaroffset]{\kern #1\overline{\kern -#1 #2}}%

\def\besiii {\mbox{BESIII}\xspace}
\def\cleo   {\mbox{CLEO}\xspace}
\def\lhcb   {\mbox{LHCb}\xspace}

\def\Ppi         {\ensuremath{\pi}\xspace} 
\def\pion   {{\ensuremath{\Ppi}}\xspace}
\def\piz    {{\ensuremath{\pion^0}}\xspace}
\def\pip    {{\ensuremath{\pion^+}}\xspace}
\def\pim    {{\ensuremath{\pion^-}}\xspace}
\def\pipm   {{\ensuremath{\pion^\pm}}\xspace}
\def\pimp   {{\ensuremath{\pion^\mp}}\xspace}

\def\kaon    {{\ensuremath{\textit{K}}}\xspace}
\def\K      {{\ensuremath{\kaon}}\xspace}
\def\Kp      {{\ensuremath{\kaon^+}}\xspace}
\def\Km      {{\ensuremath{\kaon^-}}\xspace}
\def\Kpm     {{\ensuremath{\kaon^\pm}}\xspace}
\def\Kmp     {{\ensuremath{\kaon^\mp}}\xspace}
\def\KS      {{\ensuremath{K^0_{\mathrm{S}}}}\xspace}
\def\KL      {{\ensuremath{\kaon^0_{\mathrm{L}}}}\xspace}

\def\PD      {\ensuremath{D}\xspace}  
\def\Dbar    {{\ensuremath{\offsetoverline{\PD}}}\xspace}
\def\D       {{\ensuremath{D}}\xspace}
\def\Db      {\ensuremath{\offsetoverline{{D}}}\xspace}
\def\Dz      {{\ensuremath{\D^0}}\xspace}
\def\Dzb     {{\ensuremath{\Dbar{}^0}}\xspace}
\def\Dp      {{\ensuremath{\D^+}}\xspace}
\def\Dm      {{\ensuremath{\D^-}}\xspace}

\def\DpDm    {\ensuremath{\Dp {\kern -0.16em \Dm}}\xspace}
\def\Dstar   {{\ensuremath{\D^*}}\xspace}
\def\Dstarb  {{\ensuremath{\Dbar{}^*}}\xspace}
\def\Dstarz  {\DSTO\xspace}
\def\Dstarzb {\aDSTO\xspace}
\def\Dstarp  {{\ensuremath{\D^{*+}}}\xspace}
\def\Dstarm  {{\ensuremath{\D^{*-}}}\xspace}

\def\DDb     {{\ensuremath{\D \Db}}\xspace}

\def\Ppsi        {\ensuremath{\psi}\xspace}  
\def\psipp  {{\ensuremath{\Ppsi(3770)}}\xspace}

\def\DstarzDstarzb    {\ensuremath{\Dstarz {\kern -0.07em \Dstarzb}}\xspace}

\def\deltaKpi      {\ensuremath{\delta_{K\pi}^\D}\xspace}

\def\CP                {{\ensuremath{C\!P}}\xspace}
\def\rKpi                {{\ensuremath{r_{K\pi}^\D}}\xspace}
\def\rKpisq                {{\ensuremath{(r_{K\pi}^\D)^2}}\xspace}

\def\xD                {{\ensuremath{x}}\xspace}
\def\yD                {{\ensuremath{y}}\xspace}

\newcommand{\aunit}[1]{\ensuremath{\text{\,#1}}}       
    
\def\fb   {\ensuremath{\aunit{fb}}\xspace}
\def\invfb   {\ensuremath{\fb^{-1}}\xspace}
 
\newcommand{\gevgevcccc}{\ensuremath{\gev^2\!/c^4}\xspace} 
\newcommand{\gevcc}{\ensuremath{\gev\!/c^2}\xspace} 

\def\gev{\GeV}
\def\rKpisq                {{\ensuremath{(r_{K\pi}^\D)^2}}\xspace}
\def\Fplus                {{\ensuremath{F_+^{\pi\pi\piz}}}\xspace}
\def\rCosDelta                {{\ensuremath{\rKpi\cos\deltaKpi}}\xspace}
\def\rSinDelta                {{\ensuremath{\rKpi\sin\deltaKpi}}\xspace}

\newcommand{\DT}[2]{\ensuremath{\DDb\to #1\ \text{vs.}\ #2}\xspace} 

\newcommand{\boldp}{{\ensuremath{\mathbf{p}}}\xspace}
\newcommand{\overp}{{\ensuremath{\mathrm{d\boldp}}}\xspace}

\begin{document}


\title{A novel measurement of the strong-phase difference between $\Dz\to K^-\pip$ and $\Dzb\to K^-\pip$ decays using \C-even and \C-odd quantum-correlated $\DD$ pairs}

\author{
\begin{small}
\begin{center}
M.~Ablikim$^{1}$, M.~N.~Achasov$^{4,c}$, P.~Adlarson$^{77}$, X.~C.~Ai$^{82}$, R.~Aliberti$^{36}$, A.~Amoroso$^{76A,76C}$, Q.~An$^{73,59,a}$, Y.~Bai$^{58}$, O.~Bakina$^{37}$, Y.~Ban$^{47,h}$
% ...
\\
\vspace{0.2cm}
(BESIII Collaboration)\\
\vspace{0.2cm} {\it
$^{1}$ Institute of High Energy Physics, Beijing 100049, People's Republic of China\\
% ...
$^{4}$ Budker Institute of Nuclear Physics SB RAS (BINP), Novosibirsk 630090, Russia\\
% ...
$^{36}$ Johannes Gutenberg University of Mainz, Johann-Joachim-Becher-Weg 45, D-55099 Mainz, Germany\\
$^{37}$ Joint Institute for Nuclear Research, 141980 Dubna, Moscow region, Russia\\
% ...
$^{47}$ Peking University, Beijing 100871, People's Republic of China\\
% ...
$^{58}$ Southeast University, Nanjing 211100, People's Republic of China\\
$^{59}$ State Key Laboratory of Particle Detection and Electronics, Beijing 100049, Hefei 230026, People's Republic of China\\
% ...
$^{73}$ University of Science and Technology of China, Hefei 230026, People's Republic of China\\
% ...
$^{76}$ University of Turin and INFN, (A)University of Turin, I-10125, Turin, Italy; (B)University of Eastern Piedmont, I-15121, Alessandria, Italy; (C)INFN, I-10125, Turin, Italy\\
$^{77}$ Uppsala University, Box 516, SE-75120 Uppsala, Sweden\\
% ...
$^{82}$ Zhengzhou University, Zhengzhou 450001, People's Republic of China\\
\vspace{0.2cm}
$^{a}$ Deceased\\
% ...
$^{c}$ Also at the Novosibirsk State University, Novosibirsk, 630090, Russia\\
% ...
$^{h}$ Also at State Key Laboratory of Nuclear Physics and Technology, Peking University, Beijing 100871, People's Republic of China\\
% ...
}\end{center}
\vspace{0.4cm}
\end{small}
}

\date{\today}

\begin{abstract}
A novel measurement technique of strong-phase differences between the decay amplitudes of $\Dz$ and \Dzb mesons is introduced which exploits quantum-correlated \DD pairs produced by \ee collisions at energies above the \psipp production threshold, where \DD pairs are produced in both even and odd eigenstates of the charge-conjugation symmetry. Employing this technique, the first determination of a $\Dz$--\Dzb relative strong phase is reported with such data samples. The strong-phase difference between $\Dz\to \Km\pip$ and $\Dzb\to \Km\pip$ decays, \deltaKpi, is measured to be $\deltaKpi=\left(192.8^{+11.0 + 1.9}_{-12.4 -2.4}\right)^\circ$, using a dataset corresponding to an integrated luminosity of 7.13 \invfb collected at center-of-mass energies between $4.13-4.23 \GeV$ by the \besiii experiment. 
\end{abstract}

\maketitle


\section{Introduction}\label{sec:introduction}
It has long been recognized that quantum-correlated \DD pairs produced in \ee collisions at the charm threshold can be exploited to determine decay properties that are necessary inputs for measurements of \CP violation and charm mixing~\cite{LHCbBinFlip,LHCbKShh} (here a \D denotes a neutral charm meson that is not necessarily in a flavor eigenstate). The \cleo and \besiii collaborations have reported measurements exploiting quantum-correlated \DD pairs produced at the \psipp resonance in many decay modes, for example in the $\D\to K^-\pi^+$ decay~\cite{CLEOQC,BESIII:2014rtm,deltaKpi}. The \DD pairs are produced through the annihilation process, $\ee \to \gamma^* \to c\bar{c}$, and thus are in an odd eigenstate of the charge conjugation operator, with eigenvalue $\C=-1$, due to the intermediate virtual photon. To satisfy \C quantum-number conservation, the amplitude of the \DD pair is
\begin{align}
    \ket{\D}\ket{\Db} + \C\ket{\Db}\ket{\D},\label{eq:QCAmp}
\end{align}
and hence the decays of the two \D mesons are correlated. In \DT{Y_1}{Y_2} decays, where $Y_1$ and $Y_2$ are modes accessible to both the \Dz and \Dzb mesons, interference occurs between the two final state paths $\Dz\to Y_1, \Dzb\to Y_2$ and $\Dz\to Y_2, \Dzb\to Y_1$. The effects on the interference can be exploited to determine the hadronic parameters associated with the decay
amplitudes of $Y_1$ and $Y_2$.

The presence of quantum correlations in \DD pairs produced at energies above the charm threshold, where additional particles may be produced, has been previously proposed \cite{Xing:1996pn,Bondar:2010qs,Rama:2015pmr,Naik:2021rnv}. However, until now, correlations in this regime have never been studied experimentally. In this paper, the \DD pairs produced in the $\ee\to\DD$, $\ee\to\DSTD$ and $\ee\to\DSTDST$ processes, where each \Dstar meson decays into a \D meson and a photon or neutral pion, are demonstrated to be quantum correlated using a dataset corresponding to an integrated luminosity of 7.13\invfb collected at center-of-mass energies between $4.13$ and $4.23 \GeV$ by the \besiii experiment. Charged conjugation is implied here and throughout the paper.

At higher energies the \C-eigenvalues of the additional particles produced in the \Dstar decays dictate that of the \DD pair. It naturally follows that the \DD pair can be in a \C-even or odd eigenstate. Table~\ref{tab:decay_chains} displays the six possible \DD production mechanisms for the processes examined in this paper. The $\gamma$ and \piz particles possess the eigenvalues $\C_\gamma=-1$ and $\C_\piz=+1$, respectively, and so that of the \DD pair produced in the process, $\ee\to\DD + n\gamma + m\piz$, is given by
\begin{align}
    \C = (\C_\gamma)^{n+1} \times (\C_\piz)^{m} = (-1)^{n+1}.
\end{align}

\begin{table}[tb]
\centering
\renewcommand{\arraystretch}{1.3}
\caption{Mechanisms by which quantum-correlated \DD pairs with eigenvalue $\C$ are produced through the \mbox{$\ee\to\DD,\ \DSTD$ and $\DSTDST$} processes.}
\begin{ruledtabular}
\begin{tabular}{ lc }
  \multicolumn{1}{c}{Production mechanism}  & $\C$ \\
 \toprule
 $\ee\to\DD\phantom{\gamma\piz*}$ & $-1$ \\ 
 $\ee\to\hspace{0.3mm}\DSTD\hspace{1.2mm}\to\DD\gamma\phantom{\piz*}$ & $+1$ \\ 
 $\ee\to\hspace{0.3mm}\DSTD\hspace{1.2mm}\to\DD\piz\phantom{\gamma*}$ & $-1$ \\ 
 $\ee\to\DSTDST\to\DD\gamma\gamma\phantom{**}$ & $-1$ \\ 
 $\ee\to\DSTDST\to\DD\piz\gamma\phantom{*}$ & $+1$ \\ 
 $\ee\to\DSTDST\to\DD\piz\piz$ & $-1$ \\
\end{tabular}
\end{ruledtabular}
\label{tab:decay_chains}
\end{table}

The quantum-correlated \DD pairs reconstructed in the dataset are exploited to measure the strong-phase difference between $\Dz\to\Km\pip$ and $\Dzb\to\Km\pip$ decays, denoted \deltaKpi, and defined through the amplitude ratio,
\begin{align}
    \rKpi e^{-i\deltaKpi} = \frac{A(\Dzb\to\Km\pip)}{A(\Dz\to\Km\pip)},
\end{align}
where $\rKpi$ is the magnitude of that ratio. The strong-phase difference \deltaKpi is of particular interest, as it is an auxiliary parameter both in measurements of charm mixing and \CP violation~\cite{LHCb:2017uzt,LHCb:2024hyb}, and measurements of \CP violation in $B^\pm\to \D\Kpm$ decays where the \D subsequently decays to the $K\pi$ final state~\cite{LHCbADSGLW}. The deviation of \deltaKpi from $\pi$ also provides information on the breaking of $U$-spin symmetry in the decays of charmed hadrons~\cite{Buccella:1994nf,Browder:1995ay,Gao:2006nb,Buccella:2019kpn}. The most precise measurement of this parameter with quantum-correlated \DD pairs is $\deltaKpi=\left(187.6^{+8.9 + 5.4}_{-9.7 -6.4}\right)^\circ$, where the first uncertainty is statistical and the second is systematic, performed by \besiii at the \psipp resonance using a dataset corresponding to an integrated luminosity of 2.93 \invfb~\cite{deltaKpi}. This measurement, however, is less precise than the indirect determination of $\deltaKpi=(190.2\pm2.8)^\circ$, obtained by the \lhcb collaboration from a fit to an ensemble of charm mixing and $b$-hadron decay data~\cite{GammaCombo}. Improved measurements with quantum-correlated \DD pairs are desirable to test the consistency of both determinations, and to maximize the precision with which \CP-violating and mixing parameters can be determined from external measurements of beauty and charm decays.

The decay rate for the process \DT{Y_1}{Y_2} is central to the discussion throughout the paper, and it is given by,
\begin{widetext}
\begin{align}
    \frac{\Gamma(\DT{Y_1}{Y_2})}{A_1^2 A_2^2} &= \left[r_1^2 + r_2^2 + 2\C R_1 R_2 r_1 r_2 \cos(\delta_1 - \delta_2)\right] \nonumber \\
    &-(1+\C)\yD\left[R_1 r_1 \cos\delta_1 (1+r_2^2) + R_2 r_2 \cos\delta_2 (1+r_1^2) \right] \nonumber \\
    &-(1+\C)\xD\left[R_1 r_1 \sin\delta_1 (1-r_2^2) + R_2 r_2 \sin\delta_2 (1-r_1^2) \right] 
    \nonumber\\
    &+\mathcal{O}(\xD^2, \yD^2),\label{eq:correlated_rate}
\end{align}
\end{widetext}
where \xD and \yD are the charm-mixing parameters defined in Ref.~\cite{GammaCombo}, and $A_i$, $r_i$, $R_i$ and $\delta_i$ are defined as
\begin{align}
    A_{i}^2 &= \int |A(\Dz\to Y_i)|^2 \overp, \\
    \bar{A}_i^2 &= \int |A(\Dzb\to Y_i)|^2 \overp, \\
    r_i^2 &= \frac{\bar{A}_i^2}{A_i^2}, \\
    R_i e^{i\delta_i} &= \frac{\int A(\Dz\to Y_i)A(\Dzb\to Y_i)^\ast \overp }{\sqrt{A_i^2 \bar{A}_i^2}},
\end{align}
and $\overp$ is an element of the phase space of the decay.
One important difference between the decay rates of \C-even and odd \DD pairs is that the former are affected by charm mixing at first order, whilst the dependence is quadratic for those in a \C-odd eigenstate. 

The remainder of the paper is organized as follows. First, the \besiii detector is described in Sec.~\ref{sec:data}. Then, the selection requirements are outlined in Sec.~\ref{sec:selection}, with a focus on those which are used to isolate the \DD pairs produced through each production mechanism in Table~\ref{tab:decay_chains}. Following this, Sec.~\ref{sec:QCDemo} presents a demonstration of the quantum correlations in each process using final states which are forbidden when $\C=1$ and enhanced when $\C=-1$, and vice versa. Finally, the strong-phase difference \deltaKpi is measured in Secs.~\ref{sec:DeltaKPi}-\ref{sec:extracting_dkpi}, before a combination with the results of Ref.~\cite{deltaKpi} is detailed in Sec.~\ref{sec:combination}.

\section{BESIII Detector, Simulation and Data Samples}
\label{sec:data}
The BESIII detector~\cite{BESIIIDetector} records symmetric $e^+e^-$ collisions provided by the BEPCII storage ring~\cite{BEPCII} in the center-of-mass energy range from 1.84 to 4.95~GeV, with a peak luminosity of $1.1 \times 10^{33}\;\text{cm}^{-2}\text{s}^{-1}$ achieved at $\sqrt{s} = 3.773\;\text{GeV}$. BESIII has collected large data samples in this energy region~\cite{WhitePaper}. The cylindrical core of the BESIII detector covers 93\% of the full solid angle and consists of a helium-based multilayer drift chamber~(MDC), a time-of-flight system~(TOF), and a CsI(Tl) electromagnetic calorimeter~(EMC), which are all enclosed in a superconducting solenoidal magnet providing a 1.0~T magnetic field. The solenoid is supported by an octagonal flux-return yoke with resistive plate counter muon identification modules interleaved with steel. 
The charged-particle momentum resolution at $1~{\rm GeV}/c$ is $0.5\%$, and the ${\rm d}E/{\rm d}x$ resolution is $6\%$ for electrons from Bhabha scattering. The EMC measures photon energies with a resolution of $2.5\%$ ($5\%$) at $1$~GeV in the barrel (end-cap) region. The time resolution in the plastic scintillator TOF barrel region is 68~ps, while that in the end-cap region was 110~ps. The end cap TOF system was upgraded in 2015 using multigap resistive plate chamber technology, providing a time resolution of 60~ps, which benefits $\sim 85$\% of the data used in this analysis~\cite{tof1,tof2,tof3}.

Monte Carlo (MC) simulated data samples produced with a {\sc geant4}-based~\cite{GEANT4} software package, which includes the geometric description of the BESIII detector and the detector response, are used to determine detection efficiencies and to estimate backgrounds. 
The simulation models the beam-energy spread and initial-state radiation (ISR) in the $e^+e^-$ annihilations with the generator {\sc kkmc}~\cite{KKMC}. The generator does not include the quantum correlations. Corrections are applied for this where required for background estimates. The inclusive MC sample includes the production of open-charm processes, the ISR production of vector charmonium(-like) states, and the continuum processes incorporated in {\sc kkmc}~\cite{KKMC}. All particle decays are modelled with {\sc evtgen}~\cite{EVTGEN} using branching fractions either taken from the Particle Data Group (PDG)~\cite{ParticleDataGroup:2024cfk}, when available, or otherwise estimated with {\sc lundcharm}~\cite{lundcharm1,lundcharm2}.
Final-state radiation from charged final-state particles is incorporated using the {\sc photos} package~\cite{PHOTOS}.

\section{Selection Requirements}\label{sec:selection}
The selection of candidates is divided into two stages. First, the \D-meson decays are reconstructed in six final states, $\Kpm\pimp$, $\Kp\Km$, $\pip\pim$, $\pip\pim\piz$, $\KS\piz$ and $\KS\pip\pim$. The \D-decay candidates are combined to form \DD pairs, from which a single best candidate is selected per event through the criteria described later in this section. Then, requirements are placed on the pairs to distinguish between the various \DD production mechanisms.

Charged tracks detected in the MDC are required to be within a polar angle ($\theta$) range of $\vert\!\cos\theta\vert<0.93$, where $\theta$ is defined with respect to the $z$ axis, which is the symmetry axis of the MDC. Furthermore, the distance of closest approach to the interaction point (IP) must be less than 10\,cm along the $z$ axis, $|V_{z}|$,   and less than 1\,cm in the transverse plane, $|V_{xy}|$. Particle identification~(PID) for charged tracks combines measurements of the energy deposited in the MDC~(d$E$/d$x$) and the flight time in the TOF to form likelihoods $\mathcal{L}(h)~(h=p,K,\pi)$ for each hadron $h$ hypothesis.
Tracks are identified as kaons and pions by comparing the likelihoods for the kaon and pion hypotheses, $\mathcal{L}(K)>\mathcal{L}(\pi)$ and $\mathcal{L}(\pi)>\mathcal{L}(K)$, respectively.

The $\piz$ mesons are reconstructed through the $\piz\to\gamma\gamma$ decay. Photon candidates are identified using isolated showers in the EMC. The deposited energy of each shower must be more than 25~MeV in the barrel region ($\vert\!\cos\theta\vert< 0.80$) and more than 50~MeV in the end cap region ($0.86 <\vert\!\cos\theta\vert< 0.92$). To exclude showers that originate from charged tracks, the angle subtended by the EMC shower and the position of the closest charged track at the EMC must be greater than 10 degrees as measured from the IP. To suppress electronic noise and showers unrelated to the event, the difference between the EMC time and the event start time is required to be within [0, 700]\,ns. The \piz candidates are reconstructed from two photons, where at least one produces a shower in the barrel of the EMC to provide the necessary energy resolution. The reconstructed invariant mass of the $\gamma\gamma$ pair must be $M_{\gamma\gamma}\in [115, 150] \MeVcc$. Finally, a kinematic fit constraining the two photons to the known \piz mass \cite{ParticleDataGroup:2024cfk} is performed which is required to converge with $\chi^2<50$.

Each $\KS$ candidate is reconstructed from two oppositely charged tracks satisfying $|V_{z}|<$ 20~cm. The two charged tracks are assigned as $\pi^+\pi^-$ without imposing further PID criteria. They are constrained to originate from a common vertex and are required to have an invariant mass $M_{\pip\pim}$ satisfying $|M_{\pip\pim} - m_\KS|<$ 12~$\MeVcc$, where $m_\KS$ is the known $\KS$ mass~\cite{ParticleDataGroup:2024cfk}. A secondary-vertex fit is performed to the \KS children tracks which must converge with $\chi^2<100$.

The \KS mesons used to form $\D\to\KS\piz$ candidates are subject to flight-distance significance requirements to reject background from $\D \to\pip\pim\piz$ decays. Specifically, the decay length of the \KS candidate is required to be greater than twice the vertex resolution away from the IP. The opposite requirement is placed on the $\pip\pim$ pair in the selection of $\D\to\pip\pim\piz$ candidates.

\begin{figure*}[htbp]
  \centering
    \includegraphics[width=0.9\textwidth]{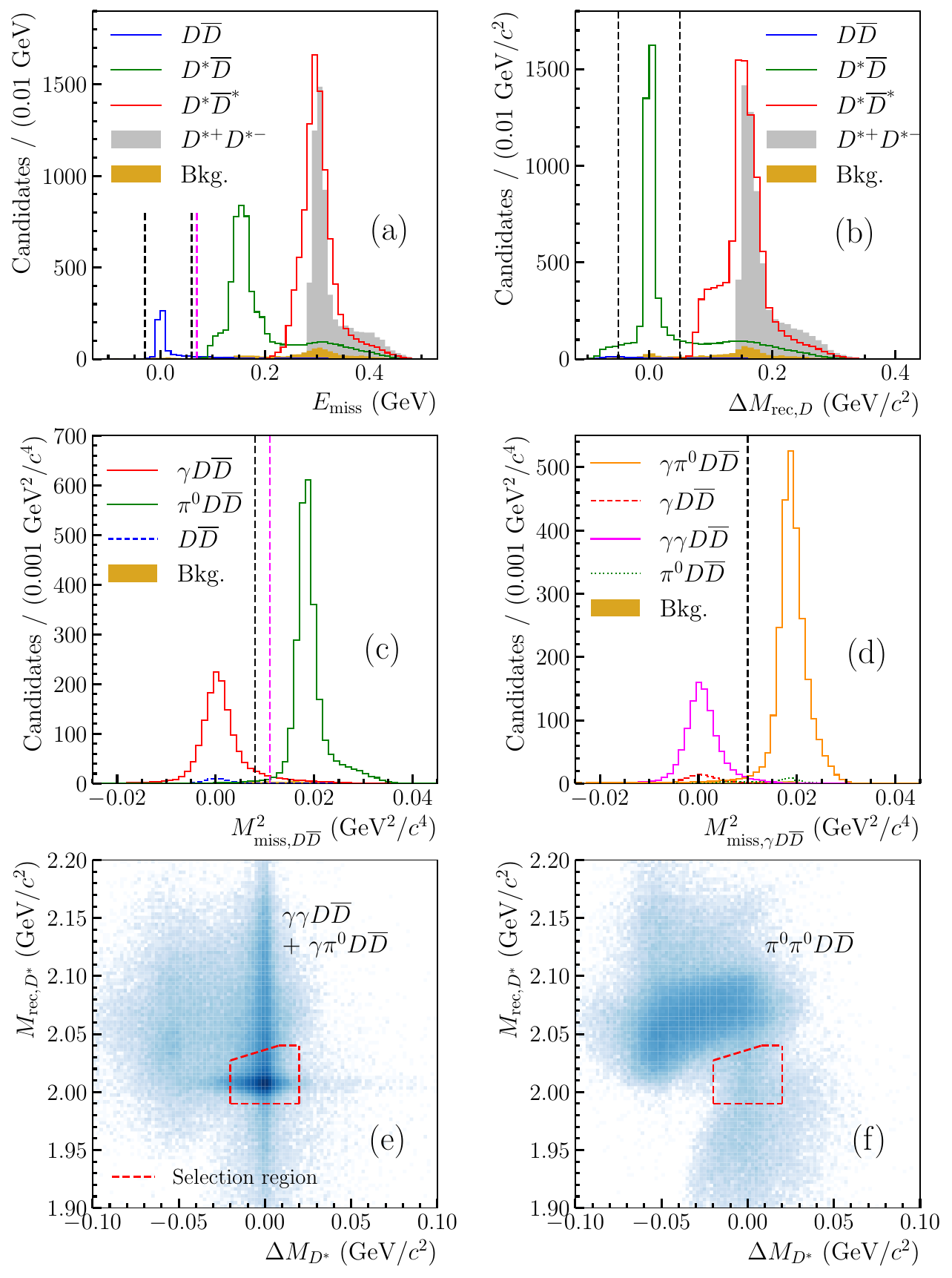}
    \caption{The (a) $\Emiss$, (b) $\MRec{D}$, (c) $\MMSq$, (d) $\MMSqGamma$ distributions of $\DD\to\Km\pip\ \mathrm{vs.}\ \Kp\pim$ candidates in MC simulation samples. The $\DeltaMDST$ vs. $\DStarMRec$ distribution of simulated \DT{\Km\pip}{\Kp\pim} decays which originate from the (e) $\DSTDST\to\gamma\gamma\DD$ and $\DSTDST\to\gamma\piz\DD$ and (f) $\DSTDST\to\piz\piz\DD$ processes are also shown. The vertical dashed lines are used to distinguish between production mechanisms and correspond to the requirements in Table~\ref{table:sels}.}
  \label{fig:SelectionPlots}
\end{figure*}

In each event the \D-decay candidates are reconstructed using the above criteria. From these, a single best \DD candidate is chosen to be that with an average reconstructed invariant mass closest to the known $\Dz$ mass~\cite{ParticleDataGroup:2024cfk}, which is subsequently used to discriminate between the different production mechanisms. The strategy is first to isolate the $\ee\to\DD$, $\ee\to\DSTD$ and $\ee\to\DSTDST$ processes, and then the possible $\Dstar$ decays. 

The isolation begins by applying a requirement on the missing energy, $\Emiss$, defined as the difference between the center-of-mass energy and that of the \DD pair,
\begin{align}
    \Emiss \equiv \Ecm -E_{\DD},
\end{align}
where $\Ecm$ is the collision energy in the center-of-mass frame and $E_{\DD}$ is the energy of the \DD pair calculated using the four-momentum, $p$, of each particle determined in an event-wide kinematic fit where all the hypothesized intermediate particles ($\KS, \piz, \Dz, \Dstarz$), are constrained to their known masses, $m$, from the PDG~\cite{ParticleDataGroup:2024cfk}. Figure~\ref{fig:SelectionPlots}(a) shows the $\Emiss$ distribution of $\DD\to\Km\pip\ \mathrm{vs.}\ \Kp\pim$ candidates in simulation. This final state is used throughout the paper as a control channel due to the large branching fraction and reconstruction efficiency of the $\Dz\to\Km\pip$ decay. Furthermore, both decays in the $\Dz\to\Kp\pim, \Dzb\to\Km\pip$ path to the final state are doubly Cabibbo suppressed compared to the alternative, $\Dz\to\Km\pip, \Dzb\to\Kp\pim$, so the quantum correlations have little impact on the decay rate. In Fig.~\ref{fig:SelectionPlots}(a), the missing energy peaks at zero for candidates that originate from the $\ee\to\DD$ process, but is larger for the $\ee\to\DSTD$ and $\ee\to\DSTDST$ processes due to the missing light neutral particles from the \Dstar decays. 

The minimum difference between the recoil mass of each \D meson and the known mass of the \Dstarz meson,
\begin{align}
    \DeltaMrec \equiv  \min\left(\left|\MRec{D_1}-m_{D^{*0}}\right|,\left|\MRec{D_2}-m_{D^{*0}}\right|\right),
\end{align}
is used to distinguish between the $\ee\to\DSTD$ and $\ee\to\DSTDST$ processes. The simulated $\DeltaMrec$ distribution for \DT{\Km\pip}{\Kp\pim} candidates that pass the $\Emiss>70 \MeV$ requirement, to remove the contribution from the $\ee\to\DD$ process, can be seen in Fig.~\ref{fig:SelectionPlots}(b). The recoil mass, $\MRec{D}$, is the invariant mass of the other particles in the event after one of the \D meson decays is reconstructed with momentum $\vec{p}_{D}$,
\begin{equation}
c^2\MRec{D} = \left[\left(\Ecm^2-\sqrt{c^2\left|\vec{p}_{D}\right|^2+m_{D^{0}}^2}\right)^2-\left|c^2\vec{p}_{D}\right|^2\right]^{\frac 12}. \label{eq:mrecD}
\end{equation}
It is possible to use $\Emiss$ to separate the $\DSTD$ processes from $\DSTDST$, however, $\MRec{\D}$ displays a better resolution because it only relies on the kinematic information of a single \D meson from each event. 

It is evident from Fig.~\ref{fig:SelectionPlots} that the requirements placed do not perfectly isolate each production mechanism. The primary cause is events with a reduced center-of-mass energy due to ISR, which produce the long tails that are clearly visible in the $\Emiss$ and $\MRec{\D}$ distributions. The cross-feed between production mechanisms is accounted for in the later steps of the analysis as discussed in Sec.~\ref{sec:QCDemo}. 

Two sources of backgrounds pass the $\Emiss$ and $\MRec{\D}$ requirements. Firstly, there are contributions from incorrectly reconstructed \D decays, which peak underneath the signal due to the \D-meson mass constraints applied when calculating $\Emiss$ and $\MRec{\D}$. These are subtracted from the samples through fits to the reconstructed invariant-masses of both \D-decays as discussed in Sec.~\ref{sec:QCDemo}. Secondly, $\Dz\Dzb$ pairs which decay to the correct final state and are produced in the $\ee\to\Dstarp\Dstarm$ process, where $\Dstarp\to\pip\Dz$, pass the requirements to isolate the \DD pairs from \DSTDST production with large rates. These are indistinguishable from signal in the aforementioned fits. However, they are significantly reduced by applying a requirement on the charged pion with the lowest momentum in each event, and through the criteria which isolate the production mechanisms that originate from the \DSTDST process discussed below.

The selection efficiency of the neutral particle from the \Dstar decay is low. Therefore, to maximize the signal yields, the \DD production mechanisms that originate from the $\DSTD$ and $\DSTDST$ processes are distinguished by partial reconstruction. For example, to discriminate between the $\Dstarz\to\gamma\Dz$ and $\Dstarz\to\piz\Dz$ decays that originate from the $\ee\to\DSTD$ process, a missing mass variable, defined as
\begin{align}
    c^4\MMSq \equiv \Emiss^2 - c^2|\vec p_{\D_1}+\vec p_{\D_2}|^2,
\end{align}
is used. Figure~\ref{fig:SelectionPlots}(c) shows the $\MMSq$ distribution in simulation.

\begin{table*}[htbp]
	\centering{}
    \renewcommand{\arraystretch}{1.75}
 \small

 	\caption{\label{table:sels}The $\DD$ pair production mechanisms of interest, their expected $\C$ eigenvalues, and the criteria applied to isolate each sample. A `$\checkmark$' in the $\DeltaMDST$ vs. $\DStarMRec$ column means that candidates are required to be in the red-dashed region displayed in  Fig.~\ref{fig:SelectionPlots}(e). }
 \begin{ruledtabular}
	\begin{tabular}{rcccccc}
		\multirow{2}{*}{Production mechanism [\;\C]} & $\Emiss$ & $\DeltaMrec$ & $\MMSq$& $\DeltaMDST$ vs. $\DStarMRec$ & $\MMSqGamma$&  $\pslow$ \\
        & ($\GeV$) & ($\gevcc$) & $(\gevcc)^2$ & region & $(\gevcc)^2$ & ($\MeV/c$) \\
		\toprule
		$\DD \;[-1]$ & $[-0.030,0.060]$ & -- &--& $\times$ & -- & -- \\
		
		$\DSTDG\;[+1]$ & $>0.070$ &$[-0.050,0.050]$&$<0.008$& $\times$ & --& --\\
		$\DSTDP\;[-1]$ & $>0.070$&$[-0.050,0.050]$&$>0.011$& $\times$ & --& --\\
		
		$\DSTDSTEven\;[+1]$ &$>0.070$&$>0.070$& --& $\checkmark$ &$<0.01$& $>100$\\
		$\DSTDSTOdd\;[-1]$ & $>0.070$&$>0.070$& --& $\checkmark$ &$>0.01$& $>100$\\
	\end{tabular}
 \end{ruledtabular}

\end{table*}

Up to this point, separation between the \DD, $\DSTDG$ and $\DSTDP$ production mechanisms has been achieved using three discriminating variables which only require reconstruction of the $\DD$ pair. However, the same strategy is much less effective in isolating the three production mechanisms that arise from \DSTDST decays due to the two additional neutral particles. Again, to maintain a high efficiency a partial reconstruction technique is employed. A single $\Dstar\to\gamma\D$ candidate from each event, chosen to be that which has a reconstructed invariant-mass closest to the $\Dstarz$ from all combinations of the two selected $\D$ mesons and all photon candidates, is used to distinguish the $\DSTDST\to\gamma\gamma\DD$ and $\DSTDST\to\piz\gamma\DD$ production mechanisms from $\DSTDST\to\piz\piz\DD$. This is achieved by selecting a region in the two-dimensional phase space defined by the difference between the reconstructed invariant mass of the \Dstar candidate and the PDG value~\cite{ParticleDataGroup:2024cfk}, denoted $\DeltaMDST$, and the recoil mass of the \Dstar candidate, $\MRec{\Dstar}$, which is defined similarly to Eq.~\eqref{eq:mrecD}.

For \DD pairs produced in $\DSTDST\to\gamma\gamma\DD$ and $\DSTDST\to\piz\gamma\DD$ decays, the corresponding distributions will peak around $\DeltaMDST\simeq 0 \gevcc$ and $\MRec{\Dstar}\simeq m_\Dstar$, as shown in Fig.~\ref{fig:SelectionPlots}(e). The long tail at $\DeltaMDST\simeq 0$ and high $\MRec{\Dstar}$ occurs due to ISR, whilst the other, shorter, tails are caused by the detector resolution. For the $\DSTDST\to\piz\piz\DD$ production mechanism, the \Dstar meson is reconstructed by combining a \D meson with a random photon, and thus the corresponding distribution will peak in a displaced region dictated by the momentum distribution of the photon, as can be seen in Fig.~\ref{fig:SelectionPlots}(f). Due to the correlation of the $\DeltaMDST$ and $\MRec{\Dstar}$ variables for $\DSTDST\to\piz\piz\DD$ events, non-rectangular selections are applied to remove misidentified $\DSTDST\to\piz\piz\DD$ events. The $\Dz\Dzb$ pairs produced in the $\Dstarp\Dstarm$ process show the same displacement, and therefore they are effectively removed by the same selections from the $\gamma\gamma\DD$ and $\gamma\piz\DD$ samples.

The $\DSTDST\to\gamma\gamma\DD$ production mechanism is distinguished from $\DSTDST\to\piz\gamma\DD$ using the missing mass after reconstructing the $\D$ and $\bar{\D}^{*}$ mesons,
\begin{align}
    c^4\MMSqGamma \equiv & (\Ecm-E_{\D\bar{D}^*})^2-c^2|\vec p_{D}+\vec p_{D^{*}}|^2.
\end{align}

Finally, two techniques can be applied to reconstruct \DD pairs produced in $\DSTDST\to\piz\piz\DD$ decays. The first proceeds by reconstructing a single $\Dstarz\to\piz\Dz$ candidate. However, due to the low momenta of the photons produced in the $\piz\to\gamma\gamma$ decay, the reconstruction efficiency is too low. The second attempt examines the candidates that do not pass the requirements placed to reconstruct the other production mechanisms originating from the $\DSTDST$ process. However, the irreducible background from the $\DSTpDSTm$ process is deemed to be too large. Therefore, in this analysis, the $\ee\to\piz\piz\DD$ production mechanism is not explicitly reconstructed. The small amount that remain in the sample are incorporated as signal alongside the $\ee\to\gamma\gamma\DD$ process where the \DD pair has the same \C-eigenvalue. Together they are referred to as $\DSTDSTOdd$.

Table~\ref{table:sels} summarizes the selection requirements applied to isolate the \DD pairs. As mentioned above, and shown in Fig.~\ref{fig:SelectionPlots}, the criteria imposed do not isolate pure samples, and there is non-negligible cross-feed between the production mechanisms. Table~\ref{table:KPiRates} displays the efficiencies determined in simulation of reconstructing a particular production mechanism given the selection requirements applied to isolate another. It should be noted that the efficiency of correctly reconstructing \DD pairs from the $\ee\to\gamma\gamma\DD$ production mechanism with the designed selection requirements is relatively high ($\sim 25\%$), but it is reduced in Table~\ref{table:KPiRates} because the definition of the $\DSTDSTOdd$ production mechanism includes the  $\ee\to\piz\piz\DSTDST$ process. 

\begin{figure*}[htbp]
  \centering
    \includegraphics[width=0.9\textwidth]{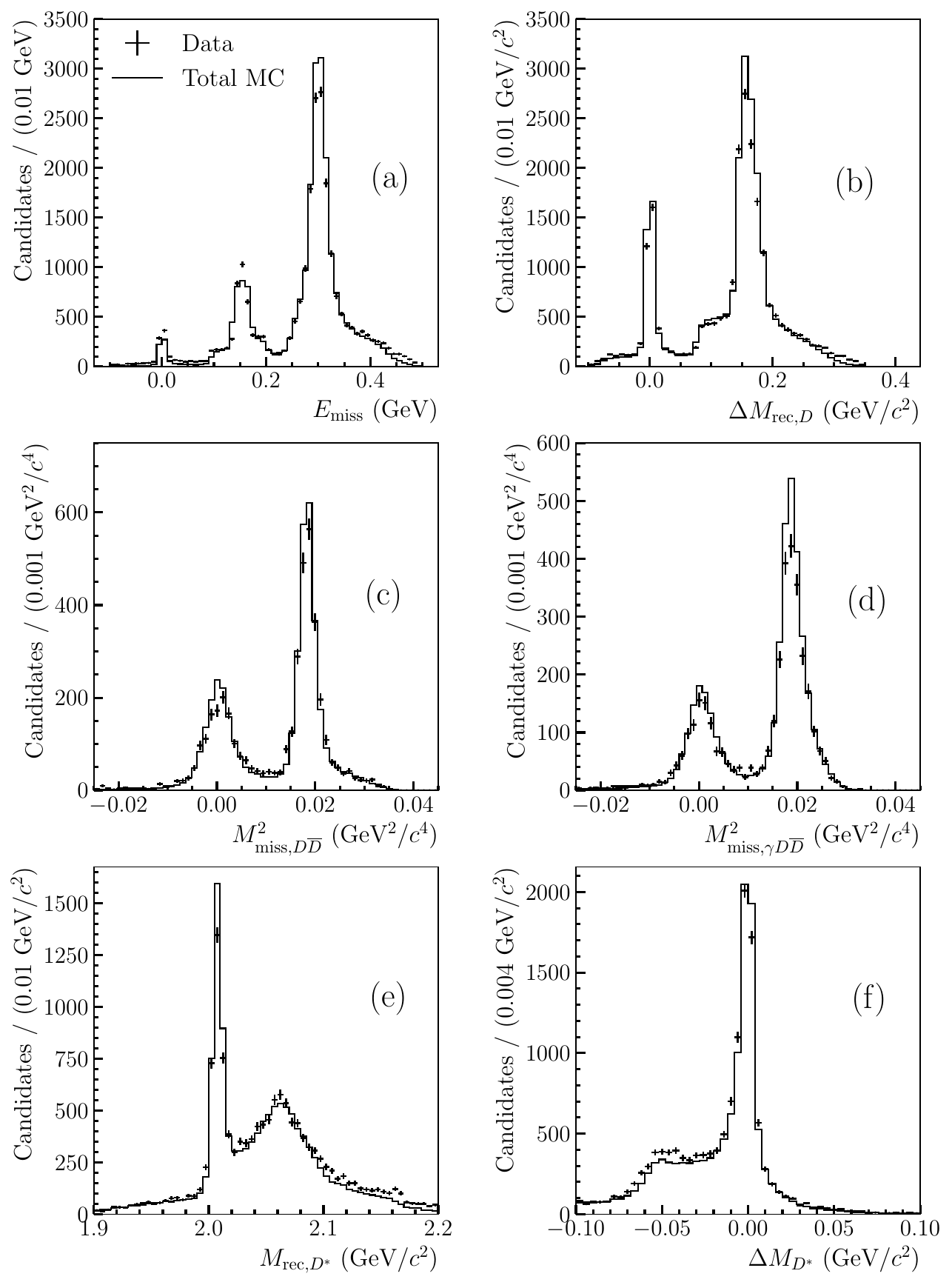}
    \caption{Comparisons between the (a) $\Emiss$, (b) $\MRec{D}$, (c) $\MMSq$, (d) $\MMSqGamma$, (e) $\DStarMRec$ and (f) $\DeltaMDST$ distributions of $\DD\to\Km\pip\ \mathrm{vs.}\ \Kp\pim$ candidates in data and MC simulation samples that contain signal and background contributions.}
  \label{fig:DataSelectionPlots}
\end{figure*}

Good agreement is found between the distributions of the selection variables in data and MC simulation samples, as shown in Fig.~\ref{fig:DataSelectionPlots}. Minor differences are observed due to the imperfect modeling of the resolution and the limited knowledge of the input cross-sections and branching fractions. The consequences of these differences are probed as systematic uncertainties, as discussed in Sec.~\ref{sec:systematics}, but are found to be small because the choice of observables leads to weak sensitivity to the related assumptions.

\begin{table*}
\caption{Efficiency, in percent, of reconstructing the $\DD\to\Km\pip\ \mathrm{vs.}\ \Kp\pim$ final state in simulation. An element with row, $i$, and column, $j$, gives the efficiency of reconstructing the production mechanism, $i$, after applying the selection requirements devised to isolate \DD pairs produced by process, $j$.}
\centering{
\renewcommand{\arraystretch}{1.7}
\begin{ruledtabular}
\begin{tabular}{ccccccc}
\multicolumn{2}{c}{}&\multicolumn{5}{c}{Identified as}     \\ 
& &  \DD & $\DSTDG$ & $\DSTDP$ & $\DSTDSTEven$ & $\DSTDSTOdd$  \\ \toprule
\multirow{4}{*}{\rotatebox[origin=c]{90}{True}}& \DD &32.45 & 2.98 & 0 & 0 & 0.07  \\ 
& $\DSTDG$ &0 & 25.44 & 1.57 & 0 & 1.94   \\ 
& $\DSTDP$ &0 & 0.40 & 30.06 & 0.55 & 0.50   \\ 
& $\DSTDSTEven$ &0 & 0 & 0 & 25.23 & 1.04 \\ 
& $\DSTDSTOdd$ &0 & 0 & 0 & 0.25 & 8.83 \\
\end{tabular}
\end{ruledtabular}
\label{table:KPiRates}
}
\end{table*}

\section{Demonstration of quantum coherence}\label{sec:QCDemo}
To demonstrate the quantum correlations in \DD pairs for the different production mechanisms, final states that are expected to be enhanced or forbidden are analyzed. 
Specifically, a \textit{coherence parameter}, $\kappa$, is determined for each production mechanism of interest by comparing the observed signal yields, after applying the relevant selection requirements, with the predictions assuming the decays of the \D and \Db mesons are independent. If the observed yields are consistent with containing contributions solely from quantum-correlated $D$ meson pairs, then $\kappa=1$, and $\kappa=0.5$ if the \DD pairs are completely uncorrelated.

Expressing the amplitude of the quantum-correlated \DD pair from Eq.~\eqref{eq:QCAmp} in terms of the \D meson \CP-even and odd eigenstates, denoted $D_\mathrm{E}$ and $D_\mathrm{O}$, respectively, gives
\begin{align}
        \ket{D_\mathrm{O}}\ket{D_\mathrm{E}} - \ket{D_\mathrm{E}}\ket{D_\mathrm{O}}\ \mathrm{when}\ \C=-1, \\
        \ket{D_\mathrm{E}}\ket{D_\mathrm{E}} - \ket{D_\mathrm{O}}\ket{D_\mathrm{O}}\ \mathrm{when}\ \C=+1,
\end{align}
from which it is evident that the decay into two final states with the same \CP is forbidden (enhanced) when $\C=-1$ ($+1$). The opposite is true for decays into a mixed-\CP final state.

\begin{figure*}[htbp]
  \centering
      \includegraphics[width=16cm]{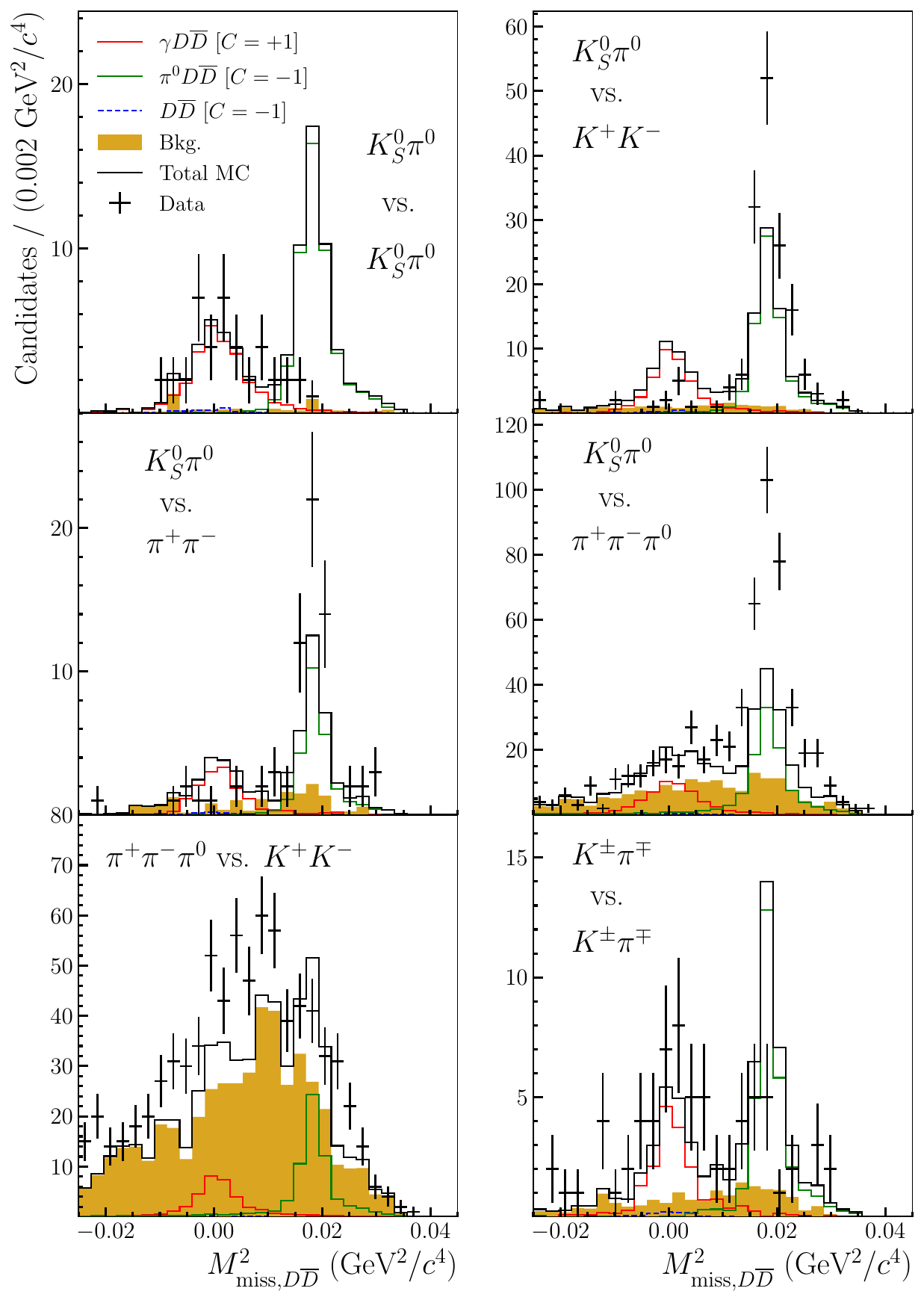}
    \caption{Comparisons between the $\MMSq$ distributions in data and MC simulation samples for each of the final states used to demonstrate the \C-coherence of the \DD pairs. The data are enhanced or suppressed relative to the MC samples, which does not account for quantum correlations, depending on the final state and the expected \C-eigenvalue of the \DD pairs that can be found in a particular region.}
    \label{fig:DD_MMiss2}
\end{figure*}

\begin{figure*}[htbp]
  \centering
      \includegraphics[width=16cm]{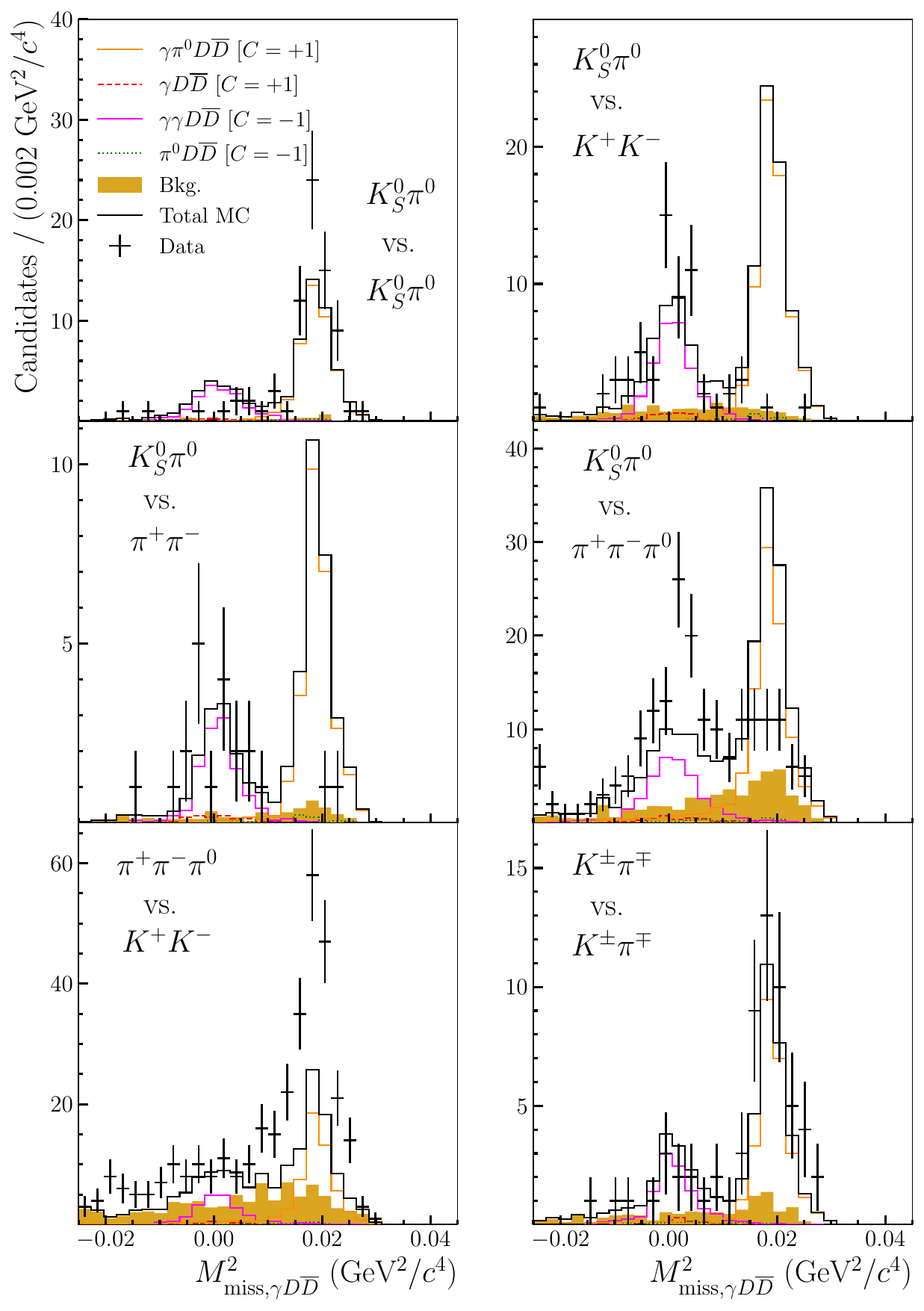}
    \caption{Comparisons between the $\MMSqGamma$ distributions in data and MC simulation samples for each of the final states used to demonstrate the \C-coherence of the \DD pairs. The data are enhanced or suppressed relative to the MC samples, which does not account for quantum correlations, depending on the final state and the expected \C-eigenvalue of the \DD pairs that can be found in a particular region.}
    \label{fig:DDg_MMiss2}
\end{figure*}

To demonstrate the quantum correlations, the \DD pairs are reconstructed in combinations of decays into the two \CP-even final states, $\D\to\Kp\Km$ and $\D\to\pip\pim$, the \CP-odd $\D\to\KS\piz$ decay and $\D\to\pip\pim\piz$. The last mode is not a \CP eigenstate but has a \CP-even fraction of $\Fplus = 0.9406 \pm0.0042$~\cite{pipipiFPlus}, and can therefore be used as a quasi \CP-even eigenstate after including a small correction to account for the \CP-odd content. Five final states consisting of two \CP eigenstates are examined, as displayed in Table~\ref{tab:CPDTags}. The other combinations of \D decays are not used, either because the purity of the sample is deemed to be too low or it provides limited additional sensitivity. The effects of the quantum correlations in each double \CP-eigenstate final state are already clear from Fig.~\ref{fig:DD_MMiss2} without a quantitative analysis. The plots compare the $\MMSq$ distributions of the candidates in data with the prediction from simulation assuming the decays of the \D and \Db mesons are independent. For the \CP-even final states, such as $\KS\piz\ \mathrm{vs.}\ \KS\piz$, the data are clearly suppressed relative to the prediction from simulation in the region $\MMSq>0.01 \gevgevcccc$ where the \C-odd \DD pairs produced through the \DSTDP production mechanism lie. The inverse behavior, where the data are enhanced relative to the simulation, is seen in the region $\MMSq<0.01 \gevgevcccc$ where the \C-even \DD pairs originating from the \DSTDG process are found. The opposite regions are enhanced and suppressed in data for the \CP-odd final states, such as $\KS\piz\ \mathrm{vs.}\ \Kp\Km$. Further examples of enhancement and suppression in the data can be seen in the $\MMSqGamma$ distributions used to distinguish between the \DSTDSTEven and \DSTDSTOdd processes, which are displayed in Fig.~\ref{fig:DDg_MMiss2}.

\begin{table}[ht!]
  \centering
    \renewcommand{\arraystretch}{1.4}
    \caption{The double \CP-eigenstate decays used throughout the measurement.}
    \label{tab:CPDTags}
    \begin{ruledtabular}
    \begin{tabular}{cc}
        $\DD\to$  & $\CP$ eigenvalue \\ \toprule
        $\pip\pim\piz\ \mathrm{vs.}\ \KS\piz$ &  $-1$\\
        \hspace{1.2mm}$\pip\pim\piz\ \mathrm{vs.}\ \Kp\Km$ &  $+1$\\
        \hspace{3.8mm}$\KS\piz\ \mathrm{vs.}\ \KS\piz$ &  $+1$\\
        \hspace{5.5mm}$\KS\piz\ \mathrm{vs.}\ \Kp\Km$ &  $-1$\\
        \hspace{4.5mm}$\KS\piz\ \mathrm{vs.}\ \pip\pim$ &  $-1$\\
    \end{tabular}
    \end{ruledtabular}
\end{table}

The $\DD\to \Kpm\pimp\ \mathrm{vs.}\ \Kpm\pimp$ decay channel also provides a test of the coherence of each production mechanism. For this final state, $r_1=r_2=\rKpi$, $\delta_1=\delta_2=\deltaKpi$ and $R_1=R_2=1$. Therefore with $\C=+1$ Eq.~\eqref{eq:correlated_rate} becomes,
\begin{align}
    4\rKpisq\left(1-\frac{(\xD\sin\deltaKpi + \yD\cos\deltaKpi)}{\rKpi}\right),
\end{align}
whilst it is suppressed to $\mathcal{O}(\xD^2, \yD^2)$ when $\C=-1$, and hence it has never been observed in quantum-correlated \DD pairs at the charm threshold. The enhancement from the mixing terms in the above equation is $\mathcal{O}(10\%)$ making the \DT{\Kmp\pipm}{\Kpm\pimp} decays from \C-even \DD pairs a promising channel for future measurements of \xD and \yD~\cite{CompanionPaper}. 

The signal yields for each final state are determined after applying the selection requirements for each production mechanism through a two-dimensional fit to the reconstructed invariant mass distributions of both $\D$ candidates. Each mass distribution contains contributions from correctly reconstructed \D decays and combinatorial background. Therefore, each fit has at least four components given by the various combinations of contributions to the mass distributions for the \D decays. In the fit, the yield of each of these components is a floating parameter. Furthermore, correctly reconstructed \D decays are modeled by a kernel density estimation (KDE)~\cite{KDE} of the distribution in simulation, convolved with a Gaussian with floating mean and width parameters. Combinatorial background components are modeled by linear polynomials  with floating slope parameters. 

\begin{figure*}[hbtp]
      \centering
      \includegraphics[width=16cm]{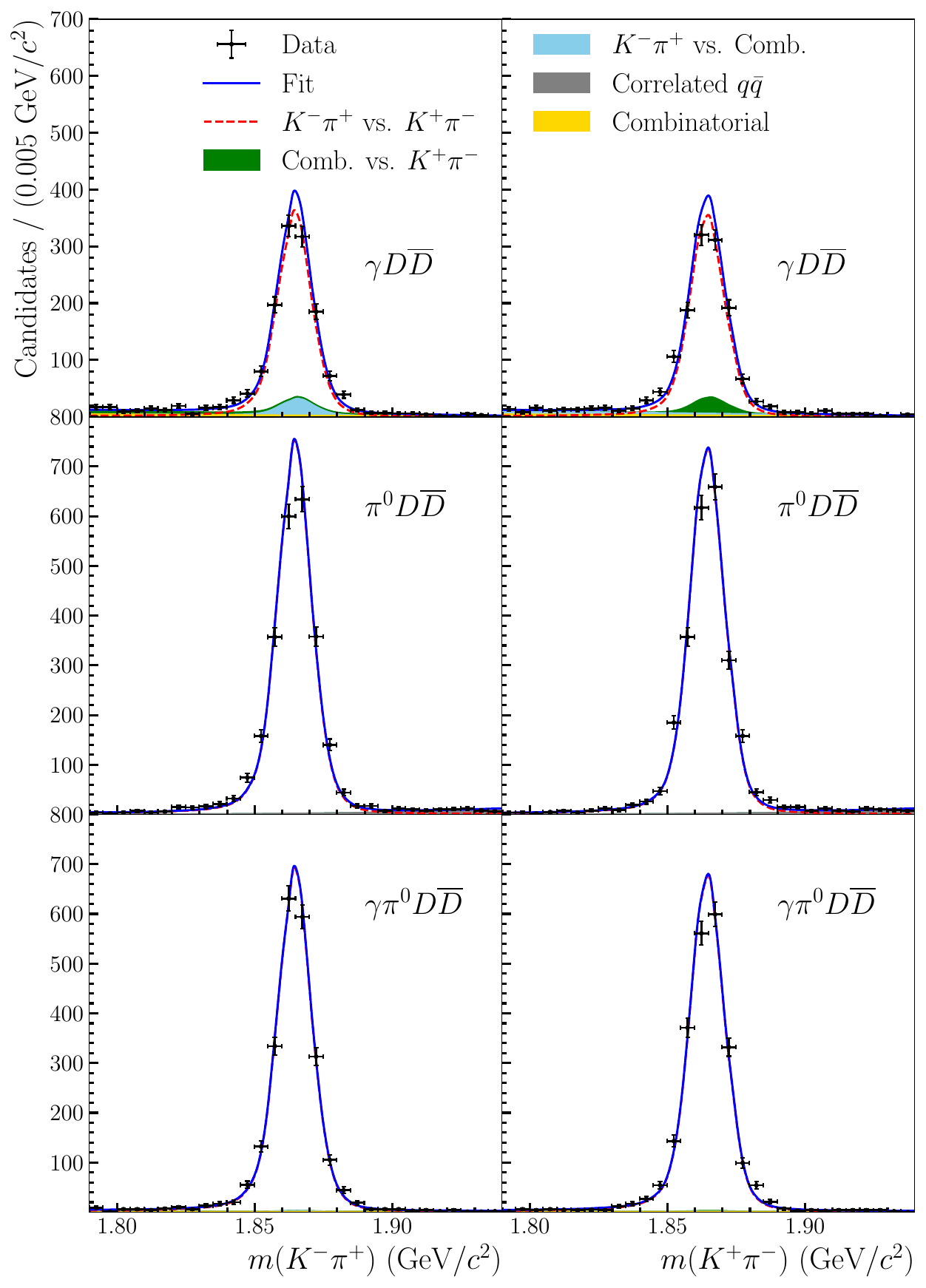}
      \caption{Projections of the fits to the $\DD\to\Km\pip\ \mathrm{vs.}\ \Kp\pim$ decays passing the requirements which isolate the (top) \DSTDG, (center) \DSTDP and (bottom) \DSTDSTEven production mechanisms.}
      \label{fig:ExampleFit}
\end{figure*}

Some fit categories include backgrounds from \ee scattering to identical final states without proceeding through an intermediate $\ee\to X\DD$ process. These are modeled by a Gaussian function in the difference of the reconstructed invariant masses with floating mean and width parameters, as the reconstructed invariant masses of these backgrounds are anti-correlated due to conservation of energy. Figure~\ref{fig:ExampleFit} displays the fits to $\DD \to K^-\pi^+\ \mathrm{vs.}\ K^+\pi^-$ candidates that pass the selection requirements to isolate the \DSTDG, \DSTDP and \DSTDSTEven production mechanisms. The latter two display the largest yields, whilst the first is chosen to show the small background contributions. 

\begin{figure*}[htbp]
  \centering
    \includegraphics[width=0.82\textwidth]{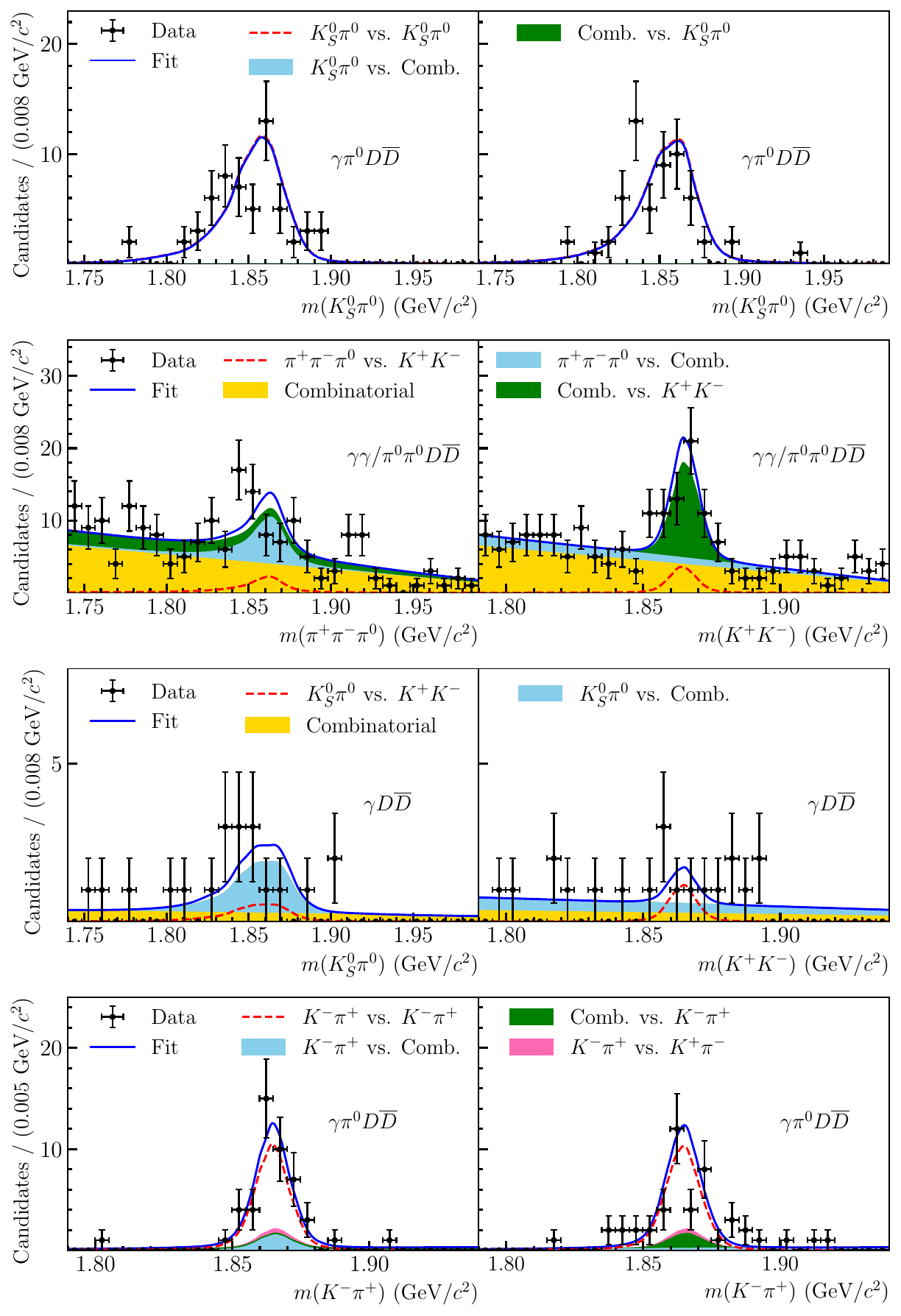}
    \caption{Projections of the fits to various final states used to demonstrate the quantum coherence of each production mechanism. The fits to the (top) \DT{\KS\piz}{\KS\piz} candidates passing the requirements for the \DSTDSTEven decays, (second from top) \DT{\pip\pim\piz}{\Kp\Km} candidates passing the requirements for the \DSTDSTOdd decays, (second from bottom) \DT{\KS\piz}{\Kp\Km} candidates passing the requirements for the \DSTDG decays and (bottom) \DT{\Km\pip}{\Km\pip} candidates passing the requirements for the \DSTDSTEven decays are shown. In each case, the left and right plots correspond to the projections of the two $\D$ decays.}
  \label{fig:example_qc_fits}
\end{figure*}

Finally, for the final states which include a $\D\to\pip\pim\piz$ decay, there is a small contribution from $\D\to\KS\piz$ decays which peak around the known $\Dz$ mass. They are not explicitly modeled in the fit and are therefore absorbed into the signal component. The fractional contribution to the signal yield from this background, $R_\KS$, is determined in simulation to be around $R_\KS \simeq 3\%$. Given the $\D\to\KS\piz$ decay is \CP odd and the $\D\to\pip\pim\piz$ decay is mostly \CP even, it is important to account for this contamination, as discussed below. The rate of the inverse scenario, where $\D\to\pip\pim\piz$ decays pass the selection criteria for the $\D\to\KS\piz$ decays, is reduced to a negligible rate by the \KS invariant mass and flight-distance significance requirements. Figure~\ref{fig:example_qc_fits} displays a sample of the fits to various final states.

\begin{table*}[hbtp]
	\centering{
    \renewcommand{\arraystretch}{2}
	\caption{Summary of the observed yields and statistical uncertainties for all double \CP-eigenstate decays in each production mechanism.}
\begin{ruledtabular}
    \begin{tabular}{rclccccc}
	\multicolumn{3}{c}{Final state}        & \DD & $\DSTDG$ & $\DSTDP$ & $\DSTDSTEven$ & $\DSTDSTOdd$\\ \toprule
	$\Km\pip$ & vs. & $\Kp\pim$ & $681.5 \pm 28.8$& $1198.0 \pm 38.8$& $2477.0 \pm 50.6$ & $2316.7 \pm 51.5$& $1146.6 \pm 37.4$\\ \hline
	$\pip\pim\piz$ & vs. & $\KS\piz$ & $80.2 ^{+ 11.0}_{-10.3} $&$23.0 ^{+ 8.0}_{-7.2} $&$272.6 ^{+ 19.2}_{-18.6}$ &$29.6 ^{+ 8.2}_{-7.3}$ &$92.9 ^{+ 11.4}_{-10.8}$\\
	$\pip\pim\piz$ & vs. & $\Kp\Km$ &    $15.4 ^{+ 9.5}_{-7.7}$ &$85.1 ^{+ 14.0}_{-12.9}$ &$21.9 ^{+ 7.7}_{-6.3}$ &$177.5 ^{+ 16.8}_{-16.0}$ &$8.9 ^{+ 6.5}_{-5.7}$\\
	$\KS\piz$ & vs. & $\KS\piz$&  $1.0 ^{+2.0}_{-1.0}$ &$30.0 ^{+ 6.2}_{-6.2}$ &$4.8 ^{+ 2.6}_{-2.6}$ &$57.7 ^{+ 8.4}_{-7.8}$ &$6.2 ^{+ 4.0}_{-3.4} $ \\
	$\KS\piz$ & vs. & $\Kp\Km$& $28.7 ^{+ 6.0}_{-5.4}$ &$2.9^{+ 3.1}_{-2.7}$ &$100.9 ^{+ 11.3}_{-10.3}$ &$1.7 ^{+ 2.0}_{-1.3}$ &$32.5 ^{+ 6.5}_{-5.8}$\\
	$\KS\piz$ & vs. & $\pip\pim$&    $10.0 ^{+ 3.8}_{-3.2}$ &$-1.6 ^{+ 2.5}_{-2.7}$ &$54.0 ^{+ 7.8}_{-7.1} $&$-0.3 ^{+ 1.4}_{-1.4}$ &$17.6 ^{+ 4.7}_{-4.1}$\\ \hline
    $\Kpm\pimp$ & vs. & $\Kpm\pimp$ & $-0.2^{+2.1}_{-1.4}$ & $18\pm 6$ & $2\pm 3$ & $35^{+7}_{-6}$ & $5\pm 3$ \\ 
	\end{tabular} \end{ruledtabular}
\label{table:FittedYields}
}
\end{table*}

The cross-feed discussed in Sec.~\ref{sec:selection} means each fitted signal yield contains contributions from \DD pairs produced through multiple processes. Corrections for the cross-feed and selection efficiencies are simultaneously applied through the matrix equation,
\begin{equation}\label{eq:unfolding}
    \vec{n}= \mathbf{A}^{-1} \vec{N},
\end{equation}
where $\vec{n}$ gives the efficiency corrected yields for each process, $\vec{N}$ contains the observed yields after applying the selection requirements to isolate each process, as determined by the fits, and the entries of the matrix $\mathbf{A}_{ij}$ give the probabilities of a \DD event from production mechanism $i$ being identified with production mechanism $j$, as determined in simulation. An example of such a matrix can be found in Table~\ref{table:KPiRates}.

\begin{figure*}[htbp]
\includegraphics[width=0.98\linewidth]{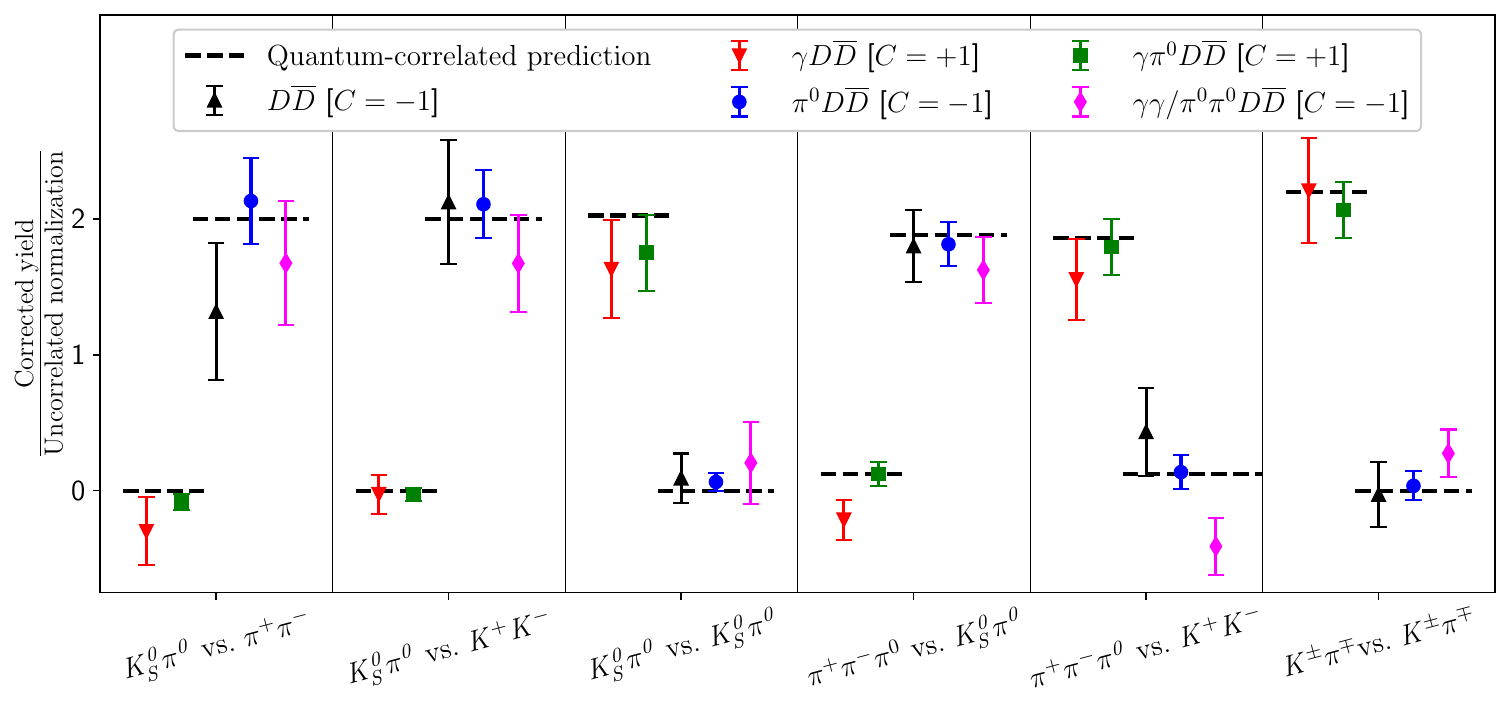}
\caption{The ratios of efficiency-corrected yields observed in data to those expected in the absence of correlations for each $\DD$ final state originating from each production mechanism. The combined statistical and systematic uncertainties are shown.}
\label{fig:QC_ratios}
\end{figure*}

The aforementioned coherence parameters describing each production mechanism are determined in a fit which compares the observed signal yields, summarized in Table~\ref{table:FittedYields}, with the predictions assuming the \D and \Db decays are independent. The predicted yields of the \DD decays to two \CP eigenstates, denoted $\DD\to Y_1\ \mathrm{vs.}\  Y_2$, are found by comparing to those measured in the $\DD\to\Km\pip\ \mathrm{vs.}\ \Kp\pim$ decay control channel,
\begin{widetext}
\begin{align}
    \vec{N}_{Y_1,Y_2} = \left(2-\delta_{Y_1}^{Y_2}\right)\frac{\mathcal B_{\Dz\to Y_1}\mathcal B_{\Dz\to Y_2}}{\mathcal B^2_{\Dz\to K^-\pi^+}} \mathbf A_{Y_1,Y_2} \mathbf \kappa \mathbf A_{K\pi,K\pi}^{-1}\vec{N}_{K\pi,K\pi}, \label{eq:uncorrelated_prediction}
\end{align}
\end{widetext}
where $(2-\delta_{Y_1}^{Y_2})$ is a combinatoric correction in terms of the Kronecker delta function and  $\mathcal B_{\Dz\to Y}$ is the $\Dz$ branching fraction to a final state $Y$. The matrix $\mathbf{\kappa}$ is defined as
\begin{align}
    \kappa_{ij} = 
        \begin{cases}
            \kappa_i\ &\textrm{when}\ i=j\ \textrm{and}\ i\ \textrm{is \C-enhanced}, \\
            1-\kappa_i\ &\textrm{when}\ i=j\ \textrm{and}\ i\ \textrm{is \C-forbidden}, \\
            0\ &\textrm{when}\ i\neq j, \\
        \end{cases}
\end{align}
such that it is diagonal with a single coherence parameter to describe each process. For final states with a $\D\to\pip\pim\piz$ decay the coherence parameters are corrected by a factor, $\Fplus/(1+R_\KS)$, to account for the \CP-odd contribution and background from $\D\to\KS\piz$ decays. 

The $\DD\to \Kpm\pimp\ \mathrm{vs.}\ \Kpm\pimp$ decays are included in the fit by examining the ratio of efficiency-corrected yields with respect to those determined in $\DD\to \Km\pip\ \mathrm{vs.}\ \Kp\pim$ decays. The predicted ratio is given by
\begin{align}\label{eq:SSRatio}
4\rKpisq f_\kappa\left(1-\frac{(\xD\sin\deltaKpi+\yD\cos\deltaKpi)}{\rKpi}\right),
\end{align}
where $f_\kappa=\kappa_i$ for processes that result in \C-even \DD pairs and $(1-\kappa_i)$ in the \C-odd scenario.

Two sources of systematic uncertainty are included in the fit. First, the knowledge of the \D decay branching fractions is accounted for by a Gaussian constraining each to the measured value from the PDG~\cite{ParticleDataGroup:2024cfk}. The second systematic uncertainty is due to efficiency differences between data and MC simulation samples in the charged-track PID, \KS and \piz reconstruction. In the fit, the selection efficiencies for each final state are multiplied by a factor which accounts for differences from data due to the three aforementioned effects. For example, in \DT{\KS\piz}{\Kp\Km} decays the efficiencies are multiplied by $(1+\Delta\epsilon_\text{PID})^2(1+\Delta\epsilon_{\KS})(1+\Delta\epsilon_\piz)$, since there are two charged tracks subject to PID, one \KS meson and one \piz meson in the final state. The PID and $\pi^0$ reconstruction efficiency differences, denoted by $\Delta\epsilon_\text{PID}$ and $\Delta\epsilon_\piz$ respectively, are Gaussian constrained to $(0\pm 1)\%$ in the fit. The $\KS$ reconstruction efficiency difference, denoted by $\Delta\epsilon_{\KS}$, is Gaussian constrained to $(0\pm 2)\%$ in the fit. The external parameters \xD, \yD, \rKpi, \deltaKpi and \Fplus are precisely known and hence they are fixed to the values determined in Refs.~\cite{GammaCombo,pipipiFPlus}. The measured coherence parameters, which are displayed in Table~\ref{tab:kappas}, are all found to be consistent with one with $\chi^2$/n.d.f. = 19.9/25, confirming the expectations. The systematic uncertainties are around 35\% of the total uncertainties shown in Table~\ref{tab:kappas}.

\begin{table}[ht!]
    \centering
    \renewcommand{\arraystretch}{1.4}
        \caption{\label{tab:kappas}Measured coherence parameters for each production mechanism.}
    \begin{ruledtabular}
    \begin{tabular}{rc  }
      Production mechanism [$\C$]   & $\kappa$    \\ \toprule 
		$\DD \;[-1]$ & $1.015\pm0.066$ \\
		
		$\DSTDG\;[+1]$ & $1.044\pm0.044$  \\
		$\DSTDP\;[-1]$ & $1.028 \pm 0.024$ \\
		
		$\DSTDSTEven\;[+1]$ & $1.027\pm0.017$ \\
		$\DSTDSTOdd\;[-1]$ &$0.963\pm0.060$ \\
    \end{tabular}
    \end{ruledtabular}
\end{table}

The effects of the quantum correlations are further evident in Fig.~\ref{fig:QC_ratios}, which shows the ratio of the efficiency-corrected yield to the prediction assuming the \DD pair is not quantum correlated for each final state and production mechanism. Once again the uncorrelated predictions are determined using the \DT{\Km\pip}{\Kp\pim} decay yields. The measured ratios are found to be in good agreement with the predictions assuming each production mechanism is in a pure \C eigenstate as computed using Eq.~\eqref{eq:correlated_rate}, which are zero for all suppressed decays and $2 + \mathcal{O}(\xD^2, \yD^2)$ for enhanced decays from \C-odd \DD pairs. Charm-mixing corrections are considered in the predictions for the enhanced decays of \C-even \DD pairs. In the decays to two \CP eigenstates the predicted ratio is $2(1 \pm 4\yD)$  where the preceding sign on \yD is negative (positive) for two \CP-even (odd) decays, and it is \mbox{$2(1 - (\yD\cos\deltaKpi + \xD\sin\deltaKpi)/\rKpi)$} for the \DT{\Kmp\pipm}{\Kmp\pipm} decays. For the final states that include a $\D\to\pip\pim\piz$ decay, a small correction is applied to the predictions to account for the \CP-odd content and the background from $\D\to\KS\piz$ decays. 

The results obtained constitute the first observation of \C-even \DD pairs. Moreover, they are the first demonstration of the quantum correlations in \DD pairs produced above the charm threshold, and opens the way for data collected at these energies to be used for measurements of the strong phases of \D mesons.

\section{Measurement of $\deltaKpi$}\label{sec:DeltaKPi}
\subsection{Overview}
A measurement of the strong-phase difference between $\Dz\to\Km\pip$ and $\Dzb\to\Km\pip$ decays, \deltaKpi, is performed. This determination assumes the expected quantum-correlated behavior of the \DD pairs from each production mechanism that is validated in Sec.~\ref{sec:QCDemo}. The strong-phase difference \deltaKpi is measured using \mbox{$\DD\to \Km\pip\ \mathrm{vs.}\ Y$} decays, where $Y$ is one of the four \CP eigenstates presented in Table~\ref{tab:CPDTags} or $\KS\pip\pim$, and is referred to as the `tag' throughout.

From Eq.~\eqref{eq:correlated_rate}, it can be shown that for the \CP tags, the quantum correlations enhance or suppress the decay rate relative to the scenario where the decays of the \D and \Db mesons are independent of each other by the factor
\begin{align}
    1 + \frac{2\C\lambda\rCosDelta - (1+\C)\yD}{1 + \rKpisq},\label{eq:cp_enhancement}
\end{align}
where terms smaller than $\mathcal{O}(\rKpi\yD)$ are neglected, and the parameter $\lambda=+1,\;-1\;\text{and }2\Fplus - 1$ for the \CP-even, odd, $\pip\pim\piz$ modes, respectively. The relative enhancement is predicted to be around $\pm 12\%$, where the preceding sign depends on the values of \C and $\lambda$. 

The \CP tags are used to measure the observable \rCosDelta. However, this alone does not allow for a single solution of \deltaKpi. Input is required from \DT{\Km\pip}{\KS\pip\pim} decays, which provide sensitivity to the observables \rCosDelta and \rSinDelta through the variation of the $\D\to\KS\pip\pim$ decay strong-phase difference across phase space. In the previous measurement of \deltaKpi performed at the \psipp resonance~\cite{deltaKpi}, \DT{\Km\pip}{\KL\pip\pim} decays were also used. However, the \KL mesons typically decay outside the \besiii detector due to their long lifetime, and so the final state is partially reconstructed. They are not used in this analysis because the strategy employed to isolate the production mechanism in Sec.~\ref{sec:selection} also relies on partial reconstruction.

In the measurement of \rCosDelta and \rSinDelta described in Sec.~\ref{sec:kspipi}, the $\D\to\KS\pip\pim$ decay phase space is divided into regions according to the ``equal-$\Delta\delta_\D$" binning scheme described in Ref.~\cite{psippKspipiPRD}. The phase space is characterized by the squared reconstructed invariant masses $m_-^2\equiv m^2(\KS\pim)$ and $m_+^2\equiv m^2(\KS\pip)$. It is divided into 16 regions which are symmetric around the line $m_-^2=m_+^2$. A region with index $j$ ($-j$) has $m_+^2 > m_-^2$ ($m_+^2 < m_-^2$) and therefore the indices range from $-8$ to 8, excluding 0. 

Inputs for the hadronic parameters of the \mbox{$\D\to\KS\pip\pim$} decay are necessary to extract the observables. They are the fraction of \mbox{$\Dz\to\KS\pip\pim$} decays that are in region $i$, denoted \Ki, and the amplitude weighted averaged cosine (sine) difference of the strong-phase difference between $\Dz \to \KS\pip\pim$ and $\Dzb \to \KS\pip\pim$ decays, denoted \ci (\si). They are defined as
\begin{widetext}
\begin{align}
    \Ki &= \int_i |A(\Dz\to\KS\pip\pim)|^2 \text{d}m_-^2\text{d}m_+^2, \\
    \Kmi &= \int_i |A(\Dzb\to\KS\pip\pim)|^2 \text{d}m_-^2\text{d}m_+^2, \\
    \ci &= \frac{1}{\sqrt{\Ki\Kmi}}\int_i |A(\Dz\to\KS\pip\pim)||A(\Dzb\to\KS\pip\pim)|\cos(\Delta\delta_D) \text{d}m_-^2\text{d}m_+^2,\label{eq:ci_def} \\
    \si &= \frac{1}{\sqrt{\Ki\Kmi}}\int_i |A(\Dz\to\KS\pip\pim)||A(\Dzb\to\KS\pip\pim)|\sin(\Delta\delta_D) \text{d}m_-^2\text{d}m_+^2\label{eq:si_def},
\end{align}
\end{widetext}
where $\Delta\delta_\D$ is the strong-phase difference between \Dz and \Dzb decays to the $\KS\pip\pim$ final state. The values used in this measurement for each are taken from Ref.~\cite{psippKspipiPRD}. With these definitions, the parameters in Eq.~\eqref{eq:correlated_rate} become $r_1\to \rKpi$, $\delta_1\to \deltaKpi$, $R_1\to R_{K\pi}=1$, $A_2^2\to \Ki$, $r_2^2 A_2^2\to \Kmi$, $R_2\cos\delta_2 \to \ci$ and $R_2\sin\delta_2 \to \si$, such that the fraction of \DT{\Km\pip}{\KS\pip\pim} decays in a region, $i$, is given by,
\begin{widetext}
\begin{align}
        Y_i &= \left[ \rKpisq \Ki + \Kmi + 2\C\rKpi\sqrt{\Ki\Kmi}\left(\ci\cos\deltaKpi + \si\sin\deltaKpi\right)\right] \nonumber \\
        &-(1+\C)\yD\left[ \rCosDelta(\Ki+\Kmi) + \ci\sqrt{\Ki\Kmi}(1+\rKpisq)\right] \nonumber \\
        &-(1+\C)\xD\left[ \rSinDelta(\Ki-\Kmi) + \si\sqrt{\Ki\Kmi}(1-\rKpisq)  \right]. \label{eq:Yi_def}
\end{align}
\end{widetext}
In the above equation, the region with index $i$ combines $\D\to\Kp\pim$ decays in region $j$ and $\D\to\Km\pip$ decays in region $-j$, where $j=i$.

\subsection{Measuring $\rCosDelta$ using \D decays to \CP-eigenstates}\label{sec:rCosDelta}
The signal yields are determined for each \CP-eigenstate tag and production mechanism using fits similar to those performed in the previous section. Alongside the five components first described in Sec.~\ref{sec:QCDemo}, additional backgrounds from \DD decays are present, which can leak into the signal regions. Therefore, fit ranges wider than those in Sec.~\ref{sec:QCDemo} are used. Consequently, an exponential function, instead of a linear polynomial, is used to describe the combinatorial background in the majority of fits, where it is found to peak towards lower masses. In the fits to $\DD\to \Km\pip\ \mathrm{vs.}\ \pip\pim\piz$ candidates which pass the selection requirements for the $\DSTDP$ and $\DSTDSTOdd$ production mechanisms, it is found that an exponential is not a good model for the combinatorial background in the reconstructed $\D\to K\pi$ invariant mass, and therefore a second order polynomial is used. The choice of background shape is considered as a systematic uncertainty.

The aforementioned backgrounds from \DD decays, the specific details of which follow, are modeled by a KDE of the distribution in simulation. Firstly,  $\D\to\Km\pip\piz$ decays can be wrongly reconstructed as $\D\to\Kp\Km$, where the $\pip$ is misidentified as a $\Kp$ and the $\piz$ is missed in the reconstruction. In the fits to candidates passing the selections to isolate \DD pairs from the $\ee\to\DD$, $\DSTDG$ and $\DSTDSTOdd$ processes the yield of this background is a floating parameter. In the others, the number passing the selections is significantly lower and therefore the yield is fixed relative to signal using the rate in simulation with a small correction factor to account for the effects of quantum correlations. In both cases the contamination is less than 10\% of the signal yield.

Background from $\D\to\Km\pip\piz$ decays is also present in the fits to candidates reconstructed as the $\D\to\pip\pim\piz$ decay, where the $\Km$ is misidentified as a $\pim$. In each fit the yield of this contamination is a floating parameter. 

Finally, an additional background is present from $\D\to\Kp\pim + \pi^0_{\Dstar}$ decays, where the $\Kp$ is misidentified as a $\pip$ and the $\piz$ originates from the $\Dstar\to\D\piz$ decay, in the fits to the $\D\to\pip\pim\piz$ candidates passing the selection criteria to isolate the $\DSTDP$ and $\DSTDSTEven$ production mechanisms. Again, in each fit the yield of this background is a floating parameter. 

\begin{figure*}[hbtp]
\centering
\begin{tabular}{c}
\includegraphics[width=0.8\textwidth]{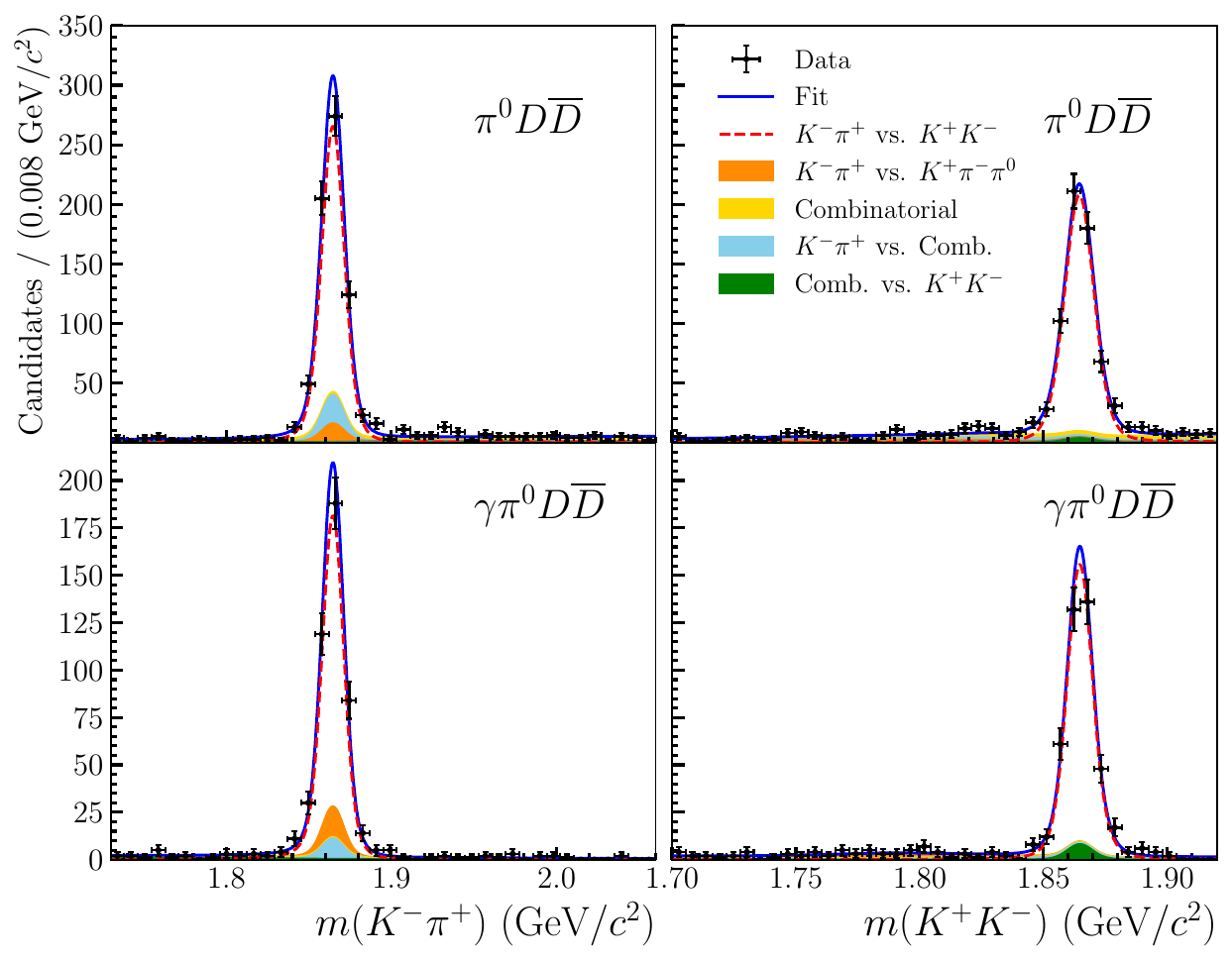} \\
\includegraphics[width=0.8\textwidth]{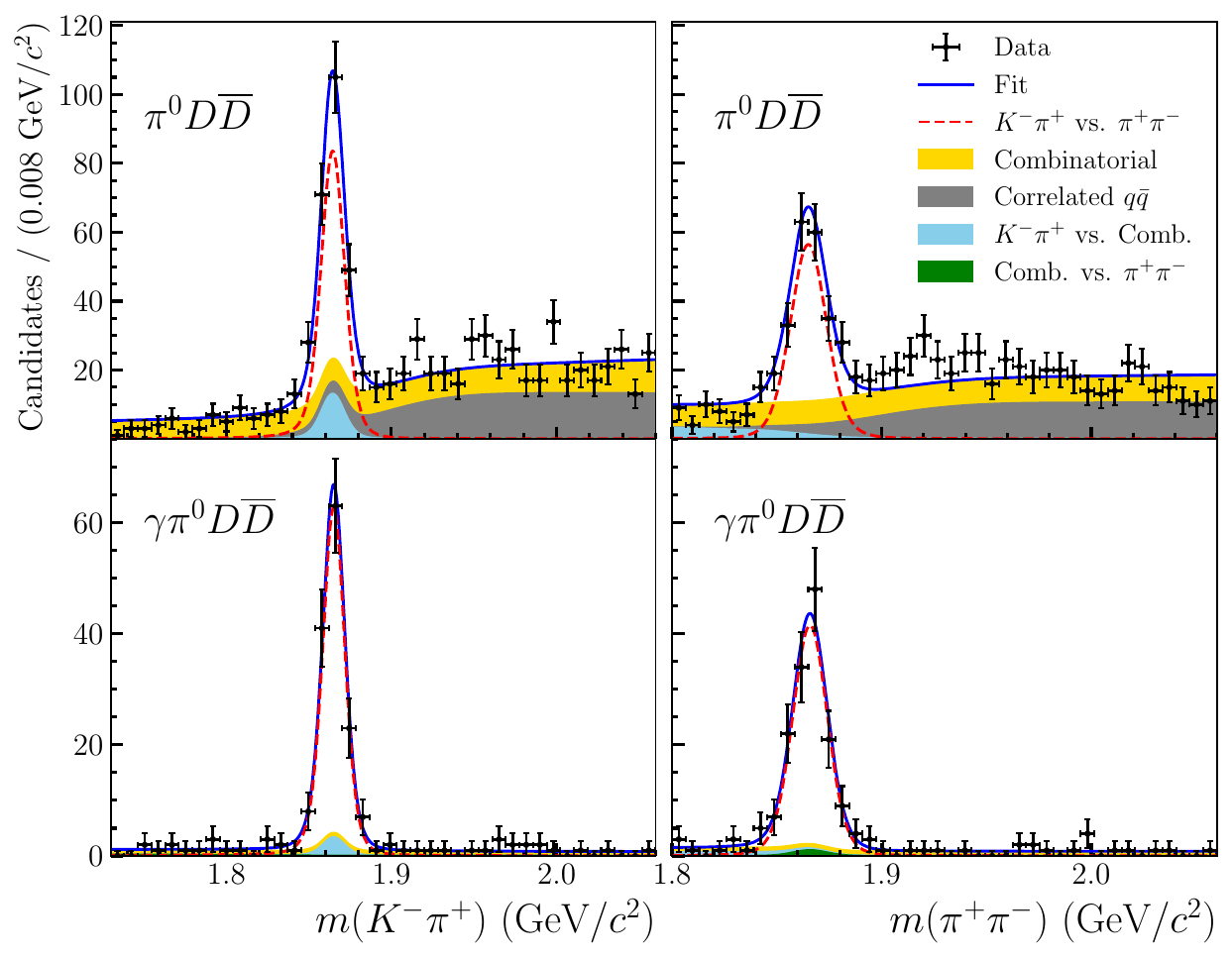} \\
\end{tabular}
\caption{Projections of the fits to (top two) \DT{\Km\pip}{\Kp\Km} (bottom two) \DT{\Km\pip}{\pip\pim} candidates passing the selection requirements which isolated the \DSTDP and \DSTDSTEven production mechanisms. The left and right columns display the invariant-mass projections for the $\Km\pip$ and \CP tag final states, respectively.}
\label{fig:kk_pipi_vs_kpi_fits}
\end{figure*}

\begin{figure*}[hbtp]
\centering
\begin{tabular}{c}
\includegraphics[width=0.8\textwidth]{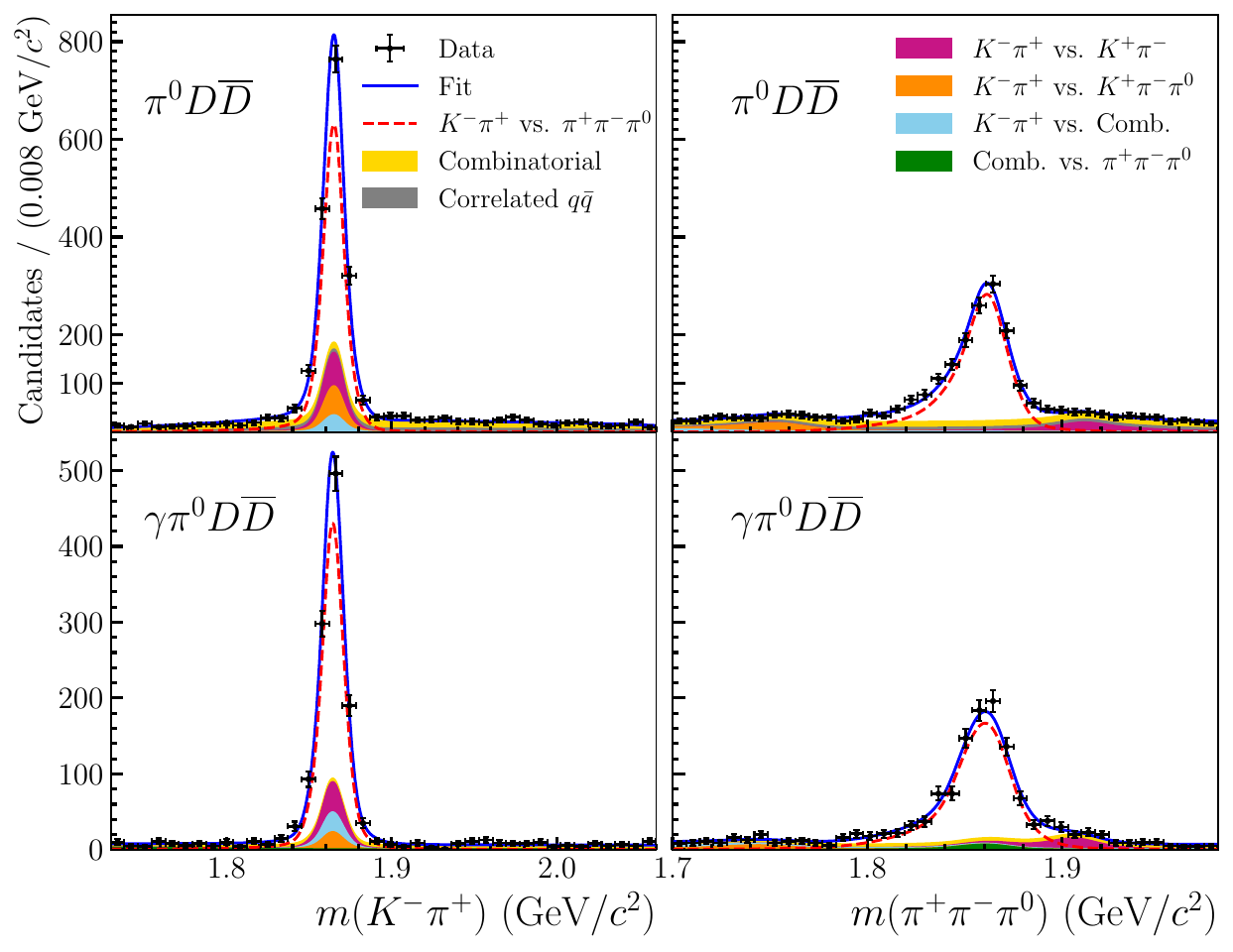} \\
\includegraphics[width=0.8\textwidth]{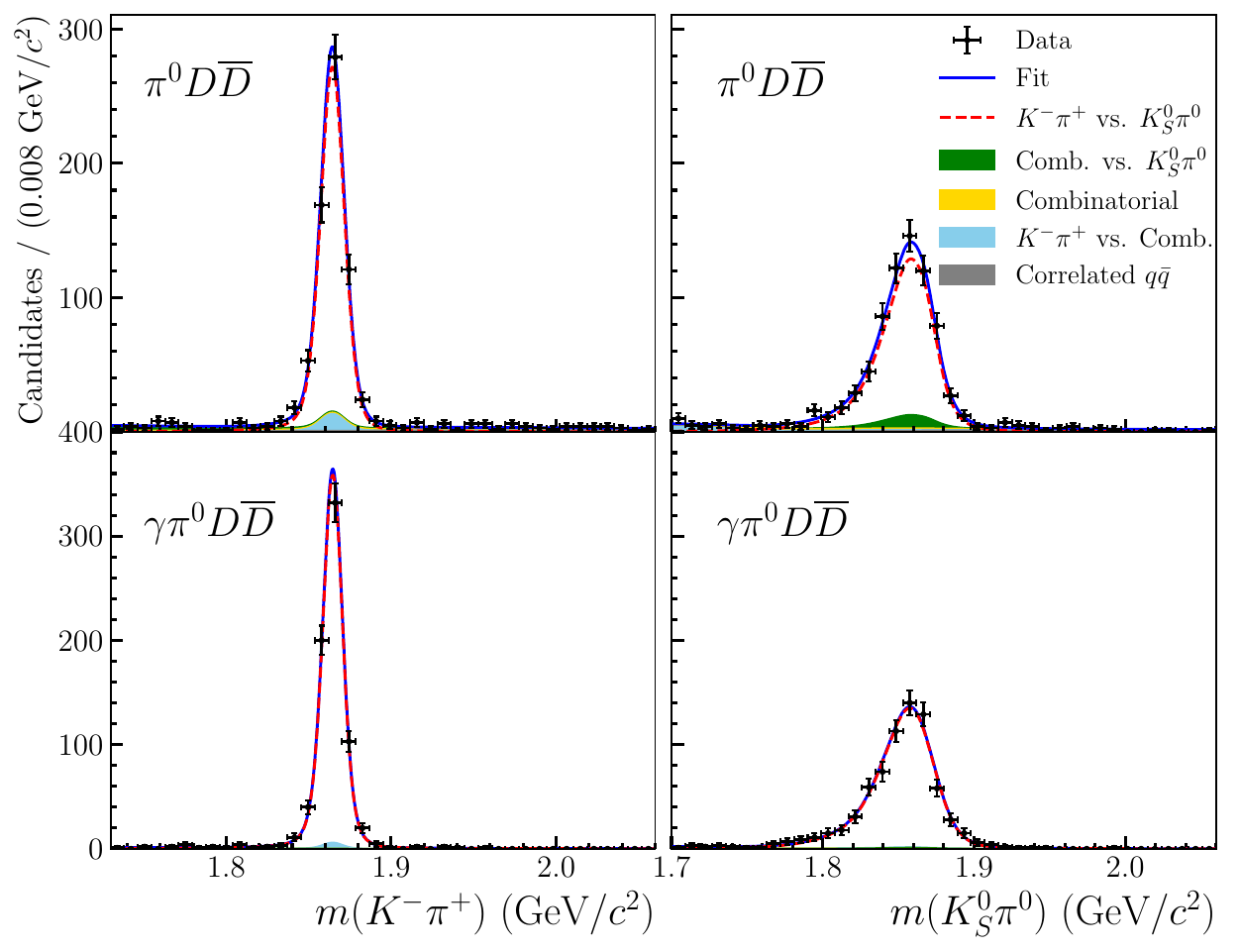} \\
\end{tabular}
\caption{Projections of the fits to (top two) \DT{\Km\pip}{\pip\pim\piz} (bottom two) \DT{\Km\pip}{\KS\piz} candidates passing the selection requirements which isolated the \DSTDP and \DSTDSTEven production mechanisms. The left and right columns display the invariant-mass projections for the $\Km\pip$ and \CP tag final states, respectively.}
\label{fig:pipipi0_kspi0_vs_kpi_fits}
\end{figure*}

Figures~\ref{fig:kk_pipi_vs_kpi_fits} and \ref{fig:pipipi0_kspi0_vs_kpi_fits} display the fits for each \CP tag of the candidates that pass the selection criteria to isolate the \DSTDP and \DSTDSTEven production mechanisms, which have the largest yields. The fitted signal yields are corrected for residual backgrounds from \DD decays to the correct final state, where the \DD pair originates from the $\ee\to\DSTpDSTm$ process, by subtracting the yield in simulation. The corrections are only applied to \DD pairs originating from the $\ee\to\DSTDST$ processes and in all cases are less than 7\% of the fitted yields. The signal yields after the corrections are displayed in Table~\ref{tab:fitted_kpi_vs_cp_yields}.
\begin{table*}[tb]
    \centering

    \renewcommand{\arraystretch}{1.8}
    \caption{Observed signal yields and statistical uncertainties in $\DD\to \Km\pip\ \mathrm{vs.}\ Y$ decays, where $Y$ is a \CP eigenstate, after applying the selection requirements to isolate each production mechanism.}
    \begin{ruledtabular}\begin{tabular}{cccccc}
         \CP eigenstate & \DD &\DSTDG & \DSTDP & \DSTDSTEven & \DSTDSTOdd \\ \toprule
        $\Kp\Km$ & {157 $\pm$ 14} & {221 $\pm$ 17} & {608 $\pm$ 27} & {395 $\pm$ 21} & {225 $\pm$ 17} \\ 
        $\pim\pim$ & {76 $\pm$ 11} &	{66 $\pm$ 10} &	{198 $\pm$ 16} & {137 $\pm$ 13} & {98 $\pm$ 11}\\ 
        $\pip\pim\piz$ & {377 $\pm$ 25} &	{578 $\pm$ 29} &	{1367 $\pm$ 41} & {915 $\pm$ 35} & {495 $\pm$ 26}\\ 
        $\KS\piz$ & {158 $\pm$ 16} &	{396 $\pm$ 21} &	{648 $\pm$ 27} & {711 $\pm$ 28} & {270 $\pm$ 18}\\
    \end{tabular} \end{ruledtabular}
    
    \label{tab:fitted_kpi_vs_cp_yields}
\end{table*}

The parameter \rCosDelta is extracted in a $\chi^2$ fit which compares the ratios of observed and predicted signal yields in the \C-even and odd processes,
\begin{align}
    \chi^2 = \sum_Y \left(\frac{R_Y^{\mathrm{obs}} - R_Y^{\mathrm{pred}}}{\delta R_Y^{\mathrm{obs}}}\right)^2,
\end{align}
where the sum runs over the \CP-eigenstate tags. The observed ratios are determined summing over the yields measured in the fits which isolate \DD pairs with \C-even, denoted E, and \C-odd, denoted O, eigenvalues,
\begin{align}
    R^{\mathrm{obs}}_Y = \frac{\sum_\mathrm{E} N^\mathrm{obs}_\mathrm{E}}{\sum_\mathrm{O} N^\mathrm{obs}_\mathrm{O}}.
\end{align}
The ratio observables are chosen to cancel all systematic uncertainties associated with the reconstruction of the final state and the total number of produced \DD pairs. The predicted yields are determined by comparing to those measured in $\DD\to \Km\pip\ \mathrm{vs.}\ \Kp\pim$ decays,
\begin{align}
    \vec{N}_{K\pi,Y} = \frac{\mathcal B_{\Dz\to \Km\pip}}{\mathcal B_{\Dz\to Y}} \mathbf A_{K\pi,Y} \mathbf f_\C^\lambda \mathbf A_{K\pi,K\pi}^{-1}\vec{N}_{K\pi,K\pi}, \label{eq:kpi_vs_cp_prediction}
\end{align}
where $\mathbf f_\C^\lambda$ is a diagonal matrix with elements given by
\begin{widetext}
 \begin{align}
    \frac{1 + \rKpisq + 2\lambda\C_\alpha\rCosDelta - \lambda(1+\C_\alpha)\yD}{1 + 2\C_\alpha\left( 2(\rCosDelta)^2 - \rKpisq \right)},
\end{align}
\end{widetext}
for a row and column of index $\alpha$. The matrix gives the predicted enhancement or suppression for each production mechanism. Each diagonal element is similar to Eq.~\eqref{eq:cp_enhancement}, but the denominator differs to account for the small effect of quantum correlations on the \DT{\Km\pip}{\Kp\pim} decay rate. It should be noted that in practice the branching fractions are not required because they cancel in the ratio.

Each efficiency matrix can be approximately divided into the contributions from the reconstruction of the final state and the isolation of each production mechanism. Hence, the predicted yields are approximately $\propto \alpha A_{K\pi, K\pi} A^{-1}_{K\pi, K\pi} = \alpha$, where $\alpha$ is a constant. Therefore, using $\DD\to \Km\pip\ \mathrm{vs.}\ \Kp\pim$ decays as a control channel is particularly beneficial because it results in the partial cancellation of systematic uncertainties associated with the efficiency of isolating each production mechanism. The cancellation is not exact due to final-state differences, for example, the resolution may vary between the \CP eigenstates with and without a \piz.

In the fit, the observable \rCosDelta is a floating parameter, and the $\DD\to \Km\pip\ \mathrm{vs.}\ \Kp\pim$ yields are Gaussian constrained to those in Table~\ref{table:FittedYields}. The remaining parameters $\rKpisq$, \yD and \Fplus are fixed to the values in Refs.~\cite{GammaCombo,pipipiFPlus}. The fit determines $\rKpi\cos\deltaKpi=-0.070\pm0.008$ with $\chi^2/\rm{n.d.f.}=2.06/3$ which corresponds to a $p$-value of 56\%. 

\begin{figure}[htbp]
    \centering{
		\includegraphics[width=0.45\textwidth]{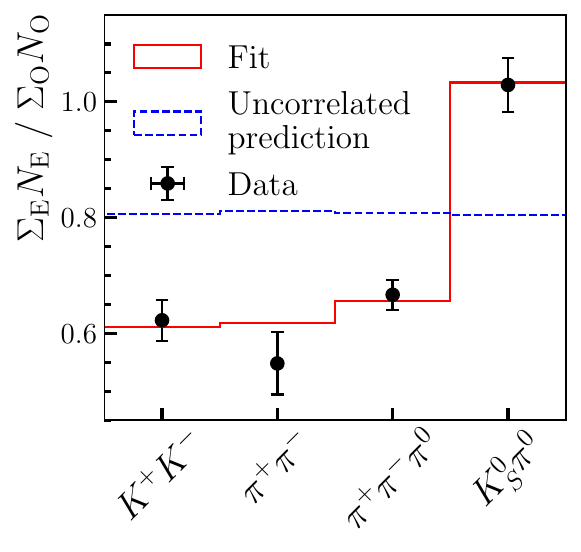} \\
    \caption{The ratios of the sum of signal yields determined after applying the requirements to isolate \C-even and \C-odd \DD pairs for each \CP tag. The result of the $\chi^2$ fit is displayed in red. }
    \label{fig:data_ratios}
 }
\end{figure}

A comparison between the observed and fitted ratios can be found in Fig.~\ref{fig:data_ratios}. The prediction in the scenario where the \D and \Db mesons decay independently is also shown, which is calculated using Eq.~\eqref{eq:kpi_vs_cp_prediction} without the enhancement matrix $\mathbf{f}_\C^\lambda$. The differences in the cross-feed efficiencies are small between the final states, and therefore the uncorrelated prediction is approximately independent of the \CP tag. From Eq.~\eqref{eq:cp_enhancement}, the final states with \CP-even tags are suppressed (enhanced) by around 12\% when they originate from \DDb pairs which are \C-odd (even). This leads to the naive expectation that the ratios for the \CP-even tags will be reduced by a factor of $(1-0.12) / (1+0.12) \simeq 0.79$ relative to the uncorrelated prediction, which is consistent with the observations. The inverse is true for the \CP-odd $\D\to\KS\piz$ tag, hence the enhancement in the ratio relative to the uncorrelated prediction.

\subsection{Measuring \rCosDelta and \rSinDelta using $\D\to\KS\pip\pim$ decays} \label{sec:kspipi}
The parameters \rCosDelta and \rSinDelta are determined by simultaneously fitting the reconstructed invariant-mass distribution of both \D mesons in 80 categories given by the region of the $\D\to \KS\pip\pim$ phase space and the hypothesized production mechanism. In addition to the five core components discussed previously, two additional background components are included in the fit. The first is from $\Dz\Dm\to [\Km\pip]_{\Dz}, [\KS\pim\piz]_\Dm$ decays, where the $\Dz\Dm$ pair originates from $\ee\to\Dstarp\Dm\to\Dz\pip\Dm$ production. These decays are only present in the categories which isolate \DD pairs originating from the $\ee\to\DSTD$ process. The selection requirements effectively remove this background in the other categories. The shape and yield of this component are fixed to those determined in simulation. The second background is from $\DD\to \Km\pip\ \mathrm{vs.}\ \KS\pip\pim$ decays, where the \DD pair originates from $\ee\to\DSTpDSTm$ production. The background is only present in the categories which isolate \DD pairs produced by the $\ee\to\DSTDST$ process. In the fit, this background is modeled with the same shape as signal and the yield is fixed to that determined in simulation.

The signal yield in each category is given by the prediction in Eq.~\eqref{eq:Yi_def} after accounting for cross-feed between the production mechanisms and migration between the phase-space regions. Therefore, in a category which isolates process, $X$, and is in phase-space region, $i$, the signal yield is given by
\begin{align}
    N_i^X = \mathbf{M}_{ij} \mathbf{A}^{|j|}_{XY} Y_j n^Y, \label{eq:kspipi_yield_param}
\end{align}
where $\mathbf{M}_{ij}$ gives the probability that a decay originally produced in region $i$ is reconstructed in region $j$. In principle, a different migration matrix could be used to describe the \DD pairs produced through each production mechanism, but they are found to be consistent in simulation so the average is used. In a similar way, the efficiency matrices in regions $j$ and $-j$ are determined to be consistent, and hence they are described by a single $\mathbf{A}^{|j|}$. The \ci, \si, \Ki, \xD and \yD parameters which are used to compute $Y_j$ are fixed to the values in Refs.~\cite{GammaCombo, psippKspipiPRD,psippKspipiPRL}. Finally, the $n^Y$ corresponds to a floating normalization parameter for process $Y$. The fit determines $\rKpi\cos\deltaKpi=-0.044 \pm 0.014$ and $\rKpi\sin\deltaKpi=-0.022 \pm 0.017$, where the uncertainties are statistical and the correlation between the observables is 3\%. The supplemental material contains a scan of all three observables as a function of the input parameters \rKpisq, \yD, \xD and \Fplus for use in combinations.

\begin{figure}[htbp]
    \centering{
		\includegraphics[width=0.48\textwidth]{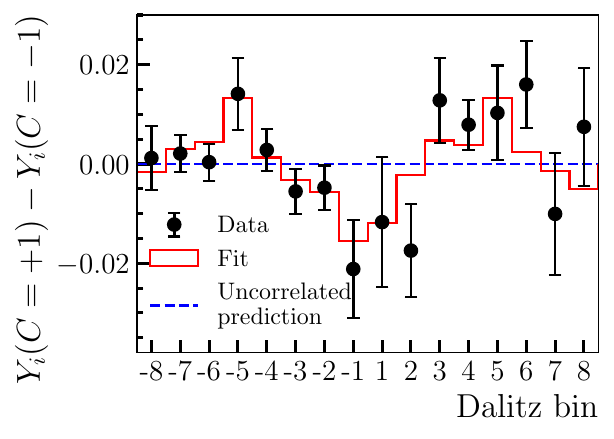} \\
    \caption{Difference between the fraction of $\DD\to \Km\pip\ \mathrm{vs.}\ \KS\pip\pim$ decays in each Dalitz region in \C-even and \C-odd production mechanisms. The differences for (solid red line) the default fit, (black points) an alternative fit where the fractional yield is a floating parameter in each category and (dashed blue line) the prediction assuming no quantum correlations are displayed.}
    \label{fig:kspipi_yields}
 }
\end{figure}

Figure~\ref{fig:kspipi_yields} displays a comparison between the difference in $Y_i$ values in the \C-even and \C-odd processes determined with the parametrization in Eq.~\eqref{eq:kspipi_yield_param}, and an alternative fit where the fraction of candidates in each phase-space region is a floating parameter for all processes. The agreement between the two is clear from the $\chi^2/\text{n.d.f.}=10.5/16$, which is significantly better than the comparison between the data and the prediction assuming no quantum correlations are present, where the $Y_i$ values are the same for \C-even and odd production mechanisms, and $\chi^2/\text{n.d.f.}=26.5/16$.

\section{Systematic uncertainties} \label{sec:systematics}
A number of systematic uncertainties associated with assumptions in the various fits are evaluated. Firstly, the uncertainty due to the knowledge of the \ci and \si inputs is determined by repeating the fit many times using an alternative set of \ci and \si generated according to their uncertainties and correlations. The standard deviations of the subsequent distributions of \rCosDelta and \rSinDelta are assigned as the systematic uncertainties and are determined to be approximately $8\times 10^{-4}$ and $27 \times 10 ^{-4}$, respectively. The uncertainty on \rSinDelta is larger than that of \rCosDelta because the \si inputs are less precise than the \ci. However, both are similar in magnitude to those reported in the previous \besiii measurement of \deltaKpi~\cite{deltaKpi}. The correlation between the observables due to this systematic uncertainty is calculated to be 8\%.

Similarly, a systematic uncertainty is included to account for the statistical uncertainties on the \Ki inputs, which are fixed in the fit, and it is estimated in a similar way to that of the \ci and \si inputs. The systematic uncertainty is found to be $11\times10^{-4}$ for \rCosDelta and $10\times10^{-4}$ for \rSinDelta with a correlation of 3\%. This is significantly reduced compared to Ref.~\cite{deltaKpi}, where this systematic uncertainty dominated. The decrease is investigated further by repeating the systematic study using predicted signal yields assuming that the \DD pairs produced through all production mechanisms are \C-even or \C-odd. The values of the observables determined in the two scenarios are found  to be $>97\%$ anti-correlated. So, varying the \Ki affects \C-even and odd \DD pairs in opposite ways, which leads to a partial cancellation and thus a reduction in the systematic uncertainty.

The cross-feed between the production mechanisms is primarily caused by missing energy due to ISR. In events with ISR the effective center-of-mass energy of the \ee pair is reduced. To model such events the simulation requires knowledge of the cross sections of the $\ee\to\DD$, $\ee\to\DSTD$ and $\ee\to\DSTDST$ processes at the reduced center-of-mass energies, where the relative uncertainties are at most around 20\% \cite{BESIII:2021yvc}. The associated systematic uncertainty is determined by repeating the fits which determine the observables using alternate efficiency matrices which are calculated in simulation where the events with ISR are reweighted by $\pm20\%$. Despite this conservative estimate the shifts in the observables are relatively small, for \rCosDelta measured with the \CP tags ($\D\to\KS\pip\pim$) it is around 10\% (4\%) of the statistical uncertainty, whilst for \rSinDelta it is found to be negligible. This systematic uncertainty is fully correlated across all observables.

A number of uncertainties are associated with the PDFs used to model different processes in the fits to determine the signal yields. As mentioned above, a choice is made for the model used to describe combinatorial background based on the fit quality. For \rCosDelta measured using the \CP tags, the associated systematic uncertainty is evaluated by performing the $\chi^2$ fit using signal yields from fits with alternative background shapes. A similar procedure is employed for the observables determined using the $\D\to\KS\pip\pim$ decay. In particular, a second-order polynomial is used for fits with combinatorial backgrounds that peak towards low masses, and a first-order polynomial is used to describe backgrounds with a relatively flat distribution. The systematic uncertainty is found to be around 9\% of the statistical uncertainty for \rCosDelta measured using the \CP tags, and approximately 2\% and 7\% for \rCosDelta and \rSinDelta from the $\KS\pip\pim$ tag, respectively.

The shape used to model the anti-correlated background is chosen somewhat arbitrarily, but provides a good-quality fit to the data. To estimate the associated uncertainty with the choice of shape, a similar procedure to that described in the  previous paragraph is performed using three alternative models which allow for additional freedom. They are an asymmetric Gaussian function with different left- and right-sided widths, a Crystal Ball function with an alternative description of the tails~\cite{Skwarnicki:1986xj}, and a Gaussian multiplied by an exponential to allow for a non-uniform distribution in $m_{\D_1} - m_{\D_2}$. The systematic uncertainty is determined to be less than 5\% of the statistical uncertainty for all observables.

In general, good agreement is found between the resolution of the selection variables in data and MC simulation samples, as seen in Fig.~\ref{fig:DataSelectionPlots}. Furthermore, any mismodeling will be reduced because the variables are calculated with constraints on the masses of the \D mesons. Regardless, the effect is probed using alternative efficiency matrices which are found by smearing the simulated distributions of the variables used in the selection of candidates discussed in Sec.~\ref{sec:selection}. For the \CP tags, a Gaussian with a width of 4.5 $\MeV$ (4.5 $\MeVcc$) for energy (mass) variables, conservatively chosen to show 
clear disagreement between the distributions in data and MC simulation samples, is used for each final state. This alternative choice of function leads to a negligible systematic uncertainty on \rCosDelta. For the $\D\to\KS\pip\pim$ tag, any efficiency differences between data and MC simulation samples are absorbed by the floating normalization parameters in the fit assuming the differences are constant across the phase-space. In general, this assumption is good. However, it may not hold for resolution differences between data and MC simulation samples. The associated systematic uncertainty is conservatively estimated by performing the fit to data many times using alternative efficiency matrices where the discriminating variables discussed in Sec.~\ref{sec:selection} are smeared by a random number between 0 and 3.5 \MeV\ in each phase-space region. The uncertainty is determined to be around 5\% of the statistical uncertainty for \rCosDelta and negligible for \rSinDelta.

The ratio of $\D\to\KS\piz$ decays passing the $\D\to\pip\pim\piz$ selection criteria, $R_\KS$, is determined using simulation. However, the ratio could be affected by mismodeling and quantum correlations. Therefore, the associated systematic uncertainty is evaluated by repeating the $\chi^2$ fit using a ratio determined in data. Specifically, the number of $\D\to\KS\piz$ decays in each sample is estimated through a two-dimensional fit to the reconstructed invariant mass of the $\D\to\Km\pip$ decay and the $\pip\pim$ pair, using candidates which satisfy $m_{\pip\pim}\in [0.35, 0.6] \gevcc$ and $m_{\pip\pim\piz}\in [1.8, 1.92] \gevcc$. The ratios determined for the five production mechanisms are in good agreement and the average is determined to be $(3.7 \pm 0.4)\%$. Using this ratio, the parameter \rCosDelta shifts by 5\% of the statistical uncertainty, which is assigned as the systematic uncertainty.

A closure check of the $\chi^2$ fit to extract \rCosDelta from the \CP tags is performed using signal yields which are generated using the values of \rKpi and \deltaKpi from Ref.~\cite{GammaCombo}. The fit returns the expected value of \rCosDelta with a bias of around $4\times 10^{-4}$. Given the bias is less than the least significant figure on the reported value, no correction is applied to \rCosDelta and it is included as a systematic uncertainty.

Additional studies are performed to estimate the effects of the uncertainty in the fixed ratios parameterizing the $\DD\to \Kpm\pimp\ \mathrm{vs.}\ \Kmp\pipm\piz$ decay background, knowledge of the $\ee\to\Dstarp\Dstarm$ and $\ee\to\Dstarp\Dm$ cross-sections, and the uncertainties of the fixed parameters in the fits to determine the observables, each of which is determined to be negligible. A summary of the systematic uncertainties is displayed in Table~\ref{tab:systematics}. The total systematic uncertainty on \rCosDelta from the \CP tags is approximately 19\% of the statistical uncertainty, whilst it is around 13\% and 18\% for \rCosDelta and \rSinDelta measured using the $\D\to\KS\pip\pim$ decay, respectively.

\begin{table}[ht!]
  \centering
    \renewcommand{\arraystretch}{1.4}
    \caption{Summary of the systematic uncertainties ($\times 10^{-4}$), on the observables \rCosDelta and \rSinDelta determined using \D decays to the \CP eigenstates and $\KS\pip\pim$. Entries denoted with `--' are negligible or do not apply to the relevant observable.}
    \label{tab:systematics}
    \begin{ruledtabular}
    \begin{tabular}{cccc}
    \multirow{2}{*}{Systematic} & \rCosDelta & \rCosDelta & \rSinDelta \\
     & $\D\to\CP$ & $\D\to\KS\pip\pim$ & $\D\to\KS\pip\pim$ \\
     \toprule
        \ci, \si & -- & 8 & 27 \\
        \Ki & -- & 11 & 10 \\
        ISR & 8 & 6 & -- \\
        $\D\to\KS\piz$ bkg & 9 & -- & -- \\
        Comb. bkg & 7 & 3 & 12 \\
        Anti-corr. bkg & 3 & 6 & 1 \\
        MC resolution & -- & 7 & -- \\
        Bias & 4 & -- & -- \\
        \hline
        Total & 15 & 18 & 31 \\
        \hline
        Statistical & 80 & 140 & 170 \\
    \end{tabular} \end{ruledtabular}
\end{table}

\section{Extracting \textbf{\deltaKpi}}\label{sec:extracting_dkpi}
The three measured \rCosDelta and \rSinDelta observables, summarized in Table~\ref{tab:observables}, are combined in a $\chi^2$ fit to determine \deltaKpi, which uses a covariance matrix that combines statistical and systematic uncertainties. The correlations between the observables due to the latter are shown in Table~\ref{tab:correlations}. Those for each individual systematic uncertainty are assumed to be zero unless explicitly mentioned in the previous section. 

The fit yields $\deltaKpi=\left(192.8^{+11.0 + 1.9}_{-12.4 -2.4}\right)^\circ$, where the first uncertainty is statistical and the second is systematic, when $\rKpi$ is fixed to the value determined in Ref.~\cite{GammaCombo}. This result is in good agreement with the global average $\deltaKpi=(190.2\pm2.8)^\circ$ from Ref.~\cite{GammaCombo}, and is competitive with the previous \besiii measurement \mbox{$\deltaKpi=\left(187.6^{+8.9 + 5.4}_{-9.7 -6.4}\right)^\circ$} \cite{deltaKpi}, made at the $\psipp$ resonance.

\begin{table}[h]
  \renewcommand{\arraystretch}{1.4}
  \centering
  \caption{Values of the measured observables. The first uncertainties are statistical and the second are systematic.}
  \begin{ruledtabular}
  \begin{tabular}{ccc} 
   Source & Observable & Value \\ \toprule
   $\D\to\CP$ & \rCosDelta & $-0.070 \pm 0.008 \pm 0.0015$ \\
   $\D\to\KS\pip\pim$ & \rCosDelta & $-0.044 \pm 0.014 \pm 0.0018$ \\
   $\D\to\KS\pip\pim$ & \rSinDelta &  $-0.022 \pm 0.017 \pm 0.0031$ \\
  \end{tabular} \end{ruledtabular}
  \label{tab:observables}
\end{table}

\begin{table}[h]
  \renewcommand{\arraystretch}{1.4}
  \centering
  \caption{The correlations between the systematic uncertainties on the observables. The observable \rCosDelta with the $^\dagger$ ($^\ddagger$) symbol is determined using the \CP-eigenstate ($\KS\pip\pim$) tags.}
  \begin{ruledtabular}
  \begin{tabular}{cccc} 
   & (\rCosDelta)$^{\dagger}$ & (\rCosDelta)$^{\ddagger}$ & \rSinDelta \\ \toprule
   (\rCosDelta)$^{\dagger}$ & 1 & 0.18 & 0\\
   (\rCosDelta)$^{\ddagger}$ & & 1 & 0.19 \\
   \rSinDelta & & & 1\\
  \end{tabular} \end{ruledtabular}
  \label{tab:correlations}
\end{table}

\section{\besiii combination}\label{sec:combination}
A maximum likelihood fit is performed to find the combined value of \deltaKpi from \besiii measurements using the observables presented in this paper and the study using data collected at the \psipp resonance~\cite{deltaKpi}. Two of the observables in Ref.~\cite{deltaKpi} are \rCosDelta and \rSinDelta determined by examining the phase space of the $\D\to\K_{S,L}^0\pip\pim$ decays. As such, the systematic uncertainty due to fixing the \ci and \si inputs is assumed to be completely correlated with the equivalent observables in this analysis. The systematic associated with the \Ki inputs is also shared between the analyses. However, the uncertainty is found to be significantly lower here, and the $\KL\pip\pim$ tags use different \Ki inputs, so the correlation is assumed to be zero. All other systematic uncertainties are uncorrelated. The full correlation matrix is displayed in the Appendix.

In the fit, the external inputs \rKpi, \yD and \Fplus are Gaussian constrained to the values in Refs.~\cite{GammaCombo,pipipiFPlus}. The result is $\deltaKpi=(189.2^{+6.9+3.4}_{-7.4-3.8})^\circ$, where the first uncertainty is statistical and the second is systematic.

\section{Outlook and Summary}\label{sec:summary}
The measurement presented in this paper is the first to use quantum-correlated \DD pairs produced above the charm threshold, and enables opportunities for similar studies in the future. The \DD pairs are examined through $\ee\to X\DD$ production, where $X$ is a combination of photons and neutral pions. These processes allow for \DD pairs in a \C-even eigenstate which, until now, have never been observed. All previous strong-phase measurements with quantum-correlated \DD pairs have been performed using data collected at the \psipp resonance where they are in a \C-odd eigenstate. 

The \C-even \DDb pairs have prospects beyond strong-phase studies. These events can be used for time-integrated measurements of the charm-mixing parameters, which impact decay rates at first order compared to a quadratic dependence in the \C-odd scenario. Such studies would provide important, independent tests of the results obtained by the \lhcb collaboration using time-dependent techniques. The prospects for future measurements are discussed further in Ref.~\cite{CompanionPaper}.

The use of \C-even and odd \DD pairs in the measurement has allowed for the deployment of novel techniques with systematic benefits compared to those used at the \psipp resonance. The strong-phase difference \deltaKpi is found to be $\deltaKpi=\left(192.8^{+11.0 + 1.9}_{-12.4 -2.4}\right)^\circ$, which is competitive with the previous \besiii determination~\cite{deltaKpi}. A combination is performed with the results of Ref.~\cite{deltaKpi} which finds a value of $\deltaKpi=(189.2^{+6.9+3.4}_{-7.4-3.8})^\circ$.

In summary, the quantum coherence in \DD pairs produced above the charm threshold has been demonstrated. A method to exploit these data without lowering reconstruction efficiency due to the presence of neutral particles has been developed. Methods where the \C-even and \C-odd states are simultaneously measured are shown to have significant advantages over corresponding methods using only a single state of quantum-correlation. These results open up a new avenue of future studies, making use of $\DD$ pairs produced in this energy regime, that will complement those made at the charm threshold. 

\section*{Acknowledgments}
The BESIII Collaboration thanks the staff of BEPCII (https://cstr.cn/31109.02.BEPC) and the IHEP computing center for their strong support. This work is supported in part by National Key R\&D Program of China under Contracts Nos. 2023YFA1606000, 2023YFA1606704; National Natural Science Foundation of China (NSFC) under Contracts Nos. 11635010, 11935015, 11935016, 11935018, 12025502, 12035009, 12035013, 12061131003, 12192260, 12192261, 12192262, 12192263, 12192264, 12192265, 12221005, 12225509, 12235017, 12361141819; the Chinese Academy of Sciences (CAS) Large-Scale Scientific Facility Program; CAS under Contract No. YSBR-101; 100 Talents Program of CAS; The Institute of Nuclear and Particle Physics (INPAC) and Shanghai Key Laboratory for Particle Physics and Cosmology; Agencia Nacional de Investigación y Desarrollo de Chile (ANID), Chile under Contract No. ANID PIA/APOYO AFB230003; ERC under Contract No. 758462; German Research Foundation DFG under Contract No. FOR5327; Istituto Nazionale di Fisica Nucleare, Italy; Knut and Alice Wallenberg Foundation under Contracts Nos. 2021.0174, 2021.0299; Ministry of Development of Turkey under Contract No. DPT2006K-120470; National Research Foundation of Korea under Contract No. NRF-2022R1A2C1092335; National Science and Technology fund of Mongolia; Polish National Science Centre under Contract No. 2024/53/B/ST2/00975; STFC (United Kingdom); Swedish Research Council under Contract No. 2019.04595; U. S. Department of Energy under Contract No. DE-FG02-05ER41374

\section*{Appendix: \besiii correlation matrix} \label{app:correlation_matrix}
Table~\ref{tab:total_correlations} displays the correlations between the observables in this analysis and in Ref.~\cite{deltaKpi} due to the systematic uncertainties.

\begin{table*}[h!]
\renewcommand{\arraystretch}{1.4}
  \centering
  \caption{The correlations between the systematic uncertainties on the observables in this measurement (denoted by an asterisk) and Ref.~\cite{deltaKpi}. The observables \rCosDelta with the $^\dagger$ ($^\ddagger$) symbol is determined using the \CP-eigenstate ($\KS\pip\pim$) tags.}
  \begin{ruledtabular}
  \begin{tabular}{cccccccc} 
   & (\rCosDelta)$^{*\dagger}$ & (\rCosDelta)$^{*\ddagger}$ & (\rSinDelta)$^*$ & $\mathcal{A}_{K\pi}$ & $\mathcal{A}_{K\pi}^{\pi\pi\piz}$ & \rCosDelta & \rSinDelta \\ \toprule
   (\rCosDelta)$^{*\dagger}$ & 1 & 0.18 & 0 & 0 & 0 & 0 & 0 \\
   (\rCosDelta)$^{*\ddagger}$ & & 1 & 0.19 & 0 & 0 & 0.09 & 0 \\
   (\rSinDelta)$^*$ & & & 1 & 0 & 0 & 0.34 & 0 \\
   $\mathcal{A}_{K\pi}$ & & & & 1 & 0.16 & 0 & 0 \\
   $\mathcal{A}_{K\pi}^{\pi\pi\piz}$ & & & & & 1 & 0 & 0 \\ 
   \rCosDelta & & & & & & 1 & 0 \\
   \rSinDelta & & & & & & & 1 \\
  \end{tabular} \end{ruledtabular}
  \label{tab:total_correlations}
\end{table*}
\clearpage
\bibliography{apssamp}

\providecommand{\noopsort}[1]{}\providecommand{\singleletter}[1]{#1}%
\begin{thebibliography}{37}%
\makeatletter
\providecommand \@ifxundefined [1]{%
 \@ifx{#1\undefined}
}%
\providecommand \@ifnum [1]{%
 \ifnum #1\expandafter \@firstoftwo
 \else \expandafter \@secondoftwo
 \fi
}%
\providecommand \@ifx [1]{%
 \ifx #1\expandafter \@firstoftwo
 \else \expandafter \@secondoftwo
 \fi
}%
\providecommand \natexlab [1]{#1}%
\providecommand \enquote  [1]{``#1''}%
\providecommand \bibnamefont  [1]{#1}%
\providecommand \bibfnamefont [1]{#1}%
\providecommand \citenamefont [1]{#1}%
\providecommand \href@noop [0]{\@secondoftwo}%
\providecommand \href [0]{\begingroup \@sanitize@url \@href}%
\providecommand \@href[1]{\@@startlink{#1}\@@href}%
\providecommand \@@href[1]{\endgroup#1\@@endlink}%
\providecommand \@sanitize@url [0]{\catcode `\\12\catcode `\$12\catcode `\&12\catcode `\#12\catcode `\^12\catcode `\_12\catcode `\%12\relax}%
\providecommand \@@startlink[1]{}%
\providecommand \@@endlink[0]{}%
\providecommand \url  [0]{\begingroup\@sanitize@url \@url }%
\providecommand \@url [1]{\endgroup\@href {#1}{\urlprefix }}%
\providecommand \urlprefix  [0]{URL }%
\providecommand \Eprint [0]{\href }%
\providecommand \doibase [0]{https://doi.org/}%
\providecommand \selectlanguage [0]{\@gobble}%
\providecommand \bibinfo  [0]{\@secondoftwo}%
\providecommand \bibfield  [0]{\@secondoftwo}%
\providecommand \translation [1]{[#1]}%
\providecommand \BibitemOpen [0]{}%
\providecommand \bibitemStop [0]{}%
\providecommand \bibitemNoStop [0]{.\EOS\space}%
\providecommand \EOS [0]{\spacefactor3000\relax}%
\providecommand \BibitemShut  [1]{\csname bibitem#1\endcsname}%
\let\auto@bib@innerbib\@empty
\bibitem [{\citenamefont {Aaij}\ \emph {et~al.}(2021{\natexlab{a}})\citenamefont {Aaij} \emph {et~al.}}]{LHCbBinFlip}%
  \BibitemOpen
  \bibfield  {author} {\bibinfo {author} {\bibfnamefont {R.}~\bibnamefont {Aaij}} \emph {et~al.} (\bibinfo {collaboration} {LHCb}),\ }\bibfield  {title} {\bibinfo {title} {{Observation of the mass difference between neutral charm-meson eigenstates}},\ }\href {https://doi.org/10.1103/PhysRevLett.127.111801} {\bibfield  {journal} {\bibinfo  {journal} {Phys. Rev. Lett.}\ }\textbf {\bibinfo {volume} {127}},\ \bibinfo {pages} {111801} (\bibinfo {year} {2021}{\natexlab{a}})},\ \Eprint {https://arxiv.org/abs/2106.03744} {arXiv:2106.03744 [hep-ex]} \BibitemShut {NoStop}%
\bibitem [{\citenamefont {Aaij}\ \emph {et~al.}(2021{\natexlab{b}})\citenamefont {Aaij} \emph {et~al.}}]{LHCbKShh}%
  \BibitemOpen
  \bibfield  {author} {\bibinfo {author} {\bibfnamefont {R.}~\bibnamefont {Aaij}} \emph {et~al.} (\bibinfo {collaboration} {LHCb}),\ }\bibfield  {title} {\bibinfo {title} {{Measurement of the CKM angle $\gamma$ in $B^\pm\to D K^\pm$ and $B^\pm \to D \pi^\pm$ decays with $D \to K_\mathrm S^0 h^+ h^-$}},\ }\href {https://doi.org/10.1007/JHEP02(2021)169} {\bibfield  {journal} {\bibinfo  {journal} {JHEP}\ }\textbf {\bibinfo {volume} {02}},\ \bibinfo {pages} {169}},\ \Eprint {https://arxiv.org/abs/2010.08483} {arXiv:2010.08483 [hep-ex]} \BibitemShut {NoStop}%
\bibitem [{\citenamefont {Asner}\ \emph {et~al.}(2008)\citenamefont {Asner} \emph {et~al.}}]{CLEOQC}%
  \BibitemOpen
  \bibfield  {author} {\bibinfo {author} {\bibfnamefont {D.~M.}\ \bibnamefont {Asner}} \emph {et~al.} (\bibinfo {collaboration} {CLEO}),\ }\bibfield  {title} {\bibinfo {title} {{Determination of the $D^0 \to K^{+} \pi^{-}$ relative strong phase using quantum-correlated measurements in $e^{+} e^{-} \to D^0 D^0$ bar at CLEO}},\ }\href {https://doi.org/10.1103/PhysRevD.78.012001} {\bibfield  {journal} {\bibinfo  {journal} {Phys. Rev. D}\ }\textbf {\bibinfo {volume} {78}},\ \bibinfo {pages} {012001} (\bibinfo {year} {2008})},\ \Eprint {https://arxiv.org/abs/0802.2268} {arXiv:0802.2268 [hep-ex]} \BibitemShut {NoStop}%
\bibitem [{\citenamefont {Ablikim}\ \emph {et~al.}(2014)\citenamefont {Ablikim} \emph {et~al.}}]{BESIII:2014rtm}%
  \BibitemOpen
  \bibfield  {author} {\bibinfo {author} {\bibfnamefont {M.}~\bibnamefont {Ablikim}} \emph {et~al.} (\bibinfo {collaboration} {BESIII}),\ }\bibfield  {title} {\bibinfo {title} {{Measurement of the $D\to K^-\pi^+$ strong phase difference in $\psi(3770)\to D^0\overline{D}{}^0$}},\ }\href {https://doi.org/10.1016/j.physletb.2014.05.071} {\bibfield  {journal} {\bibinfo  {journal} {Phys. Lett. B}\ }\textbf {\bibinfo {volume} {734}},\ \bibinfo {pages} {227} (\bibinfo {year} {2014})},\ \Eprint {https://arxiv.org/abs/1404.4691} {arXiv:1404.4691 [hep-ex]} \BibitemShut {NoStop}%
\bibitem [{\citenamefont {Ablikim}\ \emph {et~al.}(2022{\natexlab{a}})\citenamefont {Ablikim} \emph {et~al.}}]{deltaKpi}%
  \BibitemOpen
  \bibfield  {author} {\bibinfo {author} {\bibfnamefont {M.}~\bibnamefont {Ablikim}} \emph {et~al.} (\bibinfo {collaboration} {BESIII}),\ }\bibfield  {title} {\bibinfo {title} {{Improved measurement of the strong-phase difference $\delta _D^{K\pi }$ in quantum-correlated $D{\bar{D}}$ decays}},\ }\href {https://doi.org/10.1140/epjc/s10052-022-10872-2} {\bibfield  {journal} {\bibinfo  {journal} {Eur. Phys. J. C}\ }\textbf {\bibinfo {volume} {82}},\ \bibinfo {pages} {1009} (\bibinfo {year} {2022}{\natexlab{a}})},\ \Eprint {https://arxiv.org/abs/2208.09402} {arXiv:2208.09402 [hep-ex]} \BibitemShut {NoStop}%
\bibitem [{\citenamefont {Xing}(1997)}]{Xing:1996pn}%
  \BibitemOpen
  \bibfield  {author} {\bibinfo {author} {\bibfnamefont {Z.-z.}\ \bibnamefont {Xing}},\ }\bibfield  {title} {\bibinfo {title} {{$D^0$ - anti-$D^0$ mixing and $CP$ violation in neutral $D$ meson decays}},\ }\href {https://doi.org/10.1103/PhysRevD.55.196} {\bibfield  {journal} {\bibinfo  {journal} {Phys. Rev. D}\ }\textbf {\bibinfo {volume} {55}},\ \bibinfo {pages} {196} (\bibinfo {year} {1997})},\ \Eprint {https://arxiv.org/abs/hep-ph/9606422} {arXiv:hep-ph/9606422} \BibitemShut {NoStop}%
\bibitem [{\citenamefont {Bondar}\ \emph {et~al.}(2010)\citenamefont {Bondar}, \citenamefont {Poluektov},\ and\ \citenamefont {Vorobiev}}]{Bondar:2010qs}%
  \BibitemOpen
  \bibfield  {author} {\bibinfo {author} {\bibfnamefont {A.}~\bibnamefont {Bondar}}, \bibinfo {author} {\bibfnamefont {A.}~\bibnamefont {Poluektov}},\ and\ \bibinfo {author} {\bibfnamefont {V.}~\bibnamefont {Vorobiev}},\ }\bibfield  {title} {\bibinfo {title} {{Charm mixing in the model-independent analysis of correlated $\rm D^0 \bar{D}^0$ decays}},\ }\href {https://doi.org/10.1103/PhysRevD.82.034033} {\bibfield  {journal} {\bibinfo  {journal} {Phys. Rev. D}\ }\textbf {\bibinfo {volume} {82}},\ \bibinfo {pages} {034033} (\bibinfo {year} {2010})},\ \Eprint {https://arxiv.org/abs/1004.2350} {arXiv:1004.2350 [hep-ph]} \BibitemShut {NoStop}%
\bibitem [{\citenamefont {Rama}(2016)}]{Rama:2015pmr}%
  \BibitemOpen
  \bibfield  {author} {\bibinfo {author} {\bibfnamefont {M.}~\bibnamefont {Rama}},\ }\bibfield  {title} {\bibinfo {title} {{Measurement of strong phases, $D\overline{D} $ mixing, and CP violation using quantum correlation at charm threshold}},\ }\href {https://doi.org/10.1007/s11467-015-0489-6} {\bibfield  {journal} {\bibinfo  {journal} {Front. Phys. (Beijing)}\ }\textbf {\bibinfo {volume} {11}},\ \bibinfo {pages} {111404} (\bibinfo {year} {2016})}\BibitemShut {NoStop}%
\bibitem [{\citenamefont {Naik}(2023)}]{Naik:2021rnv}%
  \BibitemOpen
  \bibfield  {author} {\bibinfo {author} {\bibfnamefont {P.}~\bibnamefont {Naik}},\ }\bibfield  {title} {\bibinfo {title} {{Novel correlated $ {D}^0{\overline{D}}^0 $ systems for c/b physics and tests of T/CPT}},\ }\href {https://doi.org/10.1007/JHEP03(2023)038} {\bibfield  {journal} {\bibinfo  {journal} {JHEP}\ }\textbf {\bibinfo {volume} {03}},\ \bibinfo {pages} {038}},\ \Eprint {https://arxiv.org/abs/2102.07729} {arXiv:2102.07729 [hep-ph]} \BibitemShut {NoStop}%
\bibitem [{\citenamefont {Aaij}\ \emph {et~al.}(2018)\citenamefont {Aaij} \emph {et~al.}}]{LHCb:2017uzt}%
  \BibitemOpen
  \bibfield  {author} {\bibinfo {author} {\bibfnamefont {R.}~\bibnamefont {Aaij}} \emph {et~al.} (\bibinfo {collaboration} {LHCb}),\ }\bibfield  {title} {\bibinfo {title} {{Updated determination of $D^0$-$\overline{D}{}^0$ mixing and CP violation parameters with $D^0\to K^+\pi^-$ decays}},\ }\href {https://doi.org/10.1103/PhysRevD.97.031101} {\bibfield  {journal} {\bibinfo  {journal} {Phys. Rev. D}\ }\textbf {\bibinfo {volume} {97}},\ \bibinfo {pages} {031101} (\bibinfo {year} {2018})},\ \Eprint {https://arxiv.org/abs/1712.03220} {arXiv:1712.03220 [hep-ex]} \BibitemShut {NoStop}%
\bibitem [{\citenamefont {Aaij}\ \emph {et~al.}(2025)\citenamefont {Aaij} \emph {et~al.}}]{LHCb:2024hyb}%
  \BibitemOpen
  \bibfield  {author} {\bibinfo {author} {\bibfnamefont {R.}~\bibnamefont {Aaij}} \emph {et~al.} (\bibinfo {collaboration} {LHCb}),\ }\bibfield  {title} {\bibinfo {title} {{Measurement of $D^0-\bar{D}^0$ mixing and search for CP violation with $D^0\rightarrow \Kp\pim$ decays}},\ }\href {https://doi.org/10.1103/PhysRevD.111.012001} {\bibfield  {journal} {\bibinfo  {journal} {Phys. Rev. D}\ }\textbf {\bibinfo {volume} {111}},\ \bibinfo {pages} {012001} (\bibinfo {year} {2025})},\ \Eprint {https://arxiv.org/abs/2407.18001} {arXiv:2407.18001 [hep-ex]} \BibitemShut {NoStop}%
\bibitem [{\citenamefont {Aaij}\ \emph {et~al.}(2021{\natexlab{c}})\citenamefont {Aaij} \emph {et~al.}}]{LHCbADSGLW}%
  \BibitemOpen
  \bibfield  {author} {\bibinfo {author} {\bibfnamefont {R.}~\bibnamefont {Aaij}} \emph {et~al.} (\bibinfo {collaboration} {LHCb}),\ }\bibfield  {title} {\bibinfo {title} {{Measurement of CP observables in $B^\pm \to D^{(*)} K^\pm$ and $B^\pm \to D^{(*)} \pi^\pm$ decays using two-body $D$ final states}},\ }\href {https://doi.org/10.1007/JHEP04(2021)081} {\bibfield  {journal} {\bibinfo  {journal} {JHEP}\ }\textbf {\bibinfo {volume} {04}},\ \bibinfo {pages} {081}},\ \Eprint {https://arxiv.org/abs/2012.09903} {arXiv:2012.09903 [hep-ex]} \BibitemShut {NoStop}%
\bibitem [{\citenamefont {Buccella}\ \emph {et~al.}(1995)\citenamefont {Buccella}, \citenamefont {Lusignoli}, \citenamefont {Miele}, \citenamefont {Pugliese},\ and\ \citenamefont {Santorelli}}]{Buccella:1994nf}%
  \BibitemOpen
  \bibfield  {author} {\bibinfo {author} {\bibfnamefont {F.}~\bibnamefont {Buccella}}, \bibinfo {author} {\bibfnamefont {M.}~\bibnamefont {Lusignoli}}, \bibinfo {author} {\bibfnamefont {G.}~\bibnamefont {Miele}}, \bibinfo {author} {\bibfnamefont {A.}~\bibnamefont {Pugliese}},\ and\ \bibinfo {author} {\bibfnamefont {P.}~\bibnamefont {Santorelli}},\ }\bibfield  {title} {\bibinfo {title} {{Nonleptonic weak decays of charmed mesons}},\ }\href {https://doi.org/10.1103/PhysRevD.51.3478} {\bibfield  {journal} {\bibinfo  {journal} {Phys. Rev.}\ }\textbf {\bibinfo {volume} {D51}},\ \bibinfo {pages} {3478} (\bibinfo {year} {1995})},\ \Eprint {https://arxiv.org/abs/hep-ph/9411286} {arXiv:hep-ph/9411286} \BibitemShut {NoStop}%
\bibitem [{\citenamefont {Browder}\ and\ \citenamefont {Pakvasa}(1996)}]{Browder:1995ay}%
  \BibitemOpen
  \bibfield  {author} {\bibinfo {author} {\bibfnamefont {T.~E.}\ \bibnamefont {Browder}}\ and\ \bibinfo {author} {\bibfnamefont {S.}~\bibnamefont {Pakvasa}},\ }\bibfield  {title} {\bibinfo {title} {{Experimental implications of large \CP violation and final state interactions in the search for $\Dz-\Dzb$ mixing}},\ }\href {https://doi.org/10.1016/0370-2693(96)00776-9} {\bibfield  {journal} {\bibinfo  {journal} {Phys. Lett.}\ }\textbf {\bibinfo {volume} {B383}},\ \bibinfo {pages} {475} (\bibinfo {year} {1996})},\ \Eprint {https://arxiv.org/abs/hep-ph/9508362} {arXiv:hep-ph/9508362} \BibitemShut {NoStop}%
\bibitem [{\citenamefont {Gao}(2007)}]{Gao:2006nb}%
  \BibitemOpen
  \bibfield  {author} {\bibinfo {author} {\bibfnamefont {D.-N.}\ \bibnamefont {Gao}},\ }\bibfield  {title} {\bibinfo {title} {{Strong phases, asymmetries, and SU(3) symmetry breaking in $\D\to K\pi$ decays}},\ }\href {https://doi.org/10.1016/j.physletb.2006.11.069} {\bibfield  {journal} {\bibinfo  {journal} {Phys. Lett.}\ }\textbf {\bibinfo {volume} {B645}},\ \bibinfo {pages} {59} (\bibinfo {year} {2007})},\ \Eprint {https://arxiv.org/abs/hep-ph/0610389} {arXiv:hep-ph/0610389} \BibitemShut {NoStop}%
\bibitem [{\citenamefont {Buccella}\ \emph {et~al.}(2019)\citenamefont {Buccella}, \citenamefont {Paul},\ and\ \citenamefont {Santorelli}}]{Buccella:2019kpn}%
  \BibitemOpen
  \bibfield  {author} {\bibinfo {author} {\bibfnamefont {F.}~\bibnamefont {Buccella}}, \bibinfo {author} {\bibfnamefont {A.}~\bibnamefont {Paul}},\ and\ \bibinfo {author} {\bibfnamefont {P.}~\bibnamefont {Santorelli}},\ }\bibfield  {title} {\bibinfo {title} {{$SU(3)_F$ breaking through final state interactions and $CP$ asymmetries in $D \to PP$ decays}},\ }\href {https://doi.org/10.1103/PhysRevD.99.113001} {\bibfield  {journal} {\bibinfo  {journal} {Phys. Rev.}\ }\textbf {\bibinfo {volume} {D99}},\ \bibinfo {pages} {113001} (\bibinfo {year} {2019})},\ \Eprint {https://arxiv.org/abs/1902.05564} {arXiv:1902.05564 [hep-ph]} \BibitemShut {NoStop}%
\bibitem [{\citenamefont {Aaij}\ \emph {et~al.}(2024)\citenamefont {Aaij} \emph {et~al.}}]{GammaCombo}%
  \BibitemOpen
  \bibfield  {author} {\bibinfo {author} {\bibfnamefont {R.}~\bibnamefont {Aaij}} \emph {et~al.} (\bibinfo {collaboration} {LHCb}),\ }\href {https://doi.org/10.17181/CERN.8CUC.W3FT} {\bibinfo {title} {{Simultaneous determination of the CKM angle $\gamma$ and parameters related to mixing and $C\!P$ violation in the charm sector. CERN-LHCb-CONF-2024-004}}} (\bibinfo {year} {2024})\BibitemShut {NoStop}%
\bibitem [{\citenamefont {Ablikim}\ \emph {et~al.}(2010)\citenamefont {Ablikim} \emph {et~al.}}]{BESIIIDetector}%
  \BibitemOpen
  \bibfield  {author} {\bibinfo {author} {\bibfnamefont {M.}~\bibnamefont {Ablikim}} \emph {et~al.} (\bibinfo {collaboration} {BESIII}),\ }\bibfield  {title} {\bibinfo {title} {{Design and construction of the BESIII detector}},\ }\href {https://doi.org/10.1016/j.nima.2009.12.050} {\bibfield  {journal} {\bibinfo  {journal} {Nucl. Instrum. Meth. A}\ }\textbf {\bibinfo {volume} {614}},\ \bibinfo {pages} {345} (\bibinfo {year} {2010})},\ \Eprint {https://arxiv.org/abs/0911.4960} {arXiv:0911.4960 [physics.ins-det]} \BibitemShut {NoStop}%
\bibitem [{\citenamefont {Yu}\ \emph {et~al.}(2016)\citenamefont {Yu} \emph {et~al.}}]{BEPCII}%
  \BibitemOpen
  \bibfield  {author} {\bibinfo {author} {\bibfnamefont {C.}~\bibnamefont {Yu}} \emph {et~al.},\ }\bibfield  {title} {\bibinfo {title} {{BEPCII performance and beam dynamics studies on luminosity}},\ }in\ \href {https://doi.org/10.18429/JACoW-IPAC2016-TUYA01} {\emph {\bibinfo {booktitle} {{7th International Particle Accelerator Conference}}}}\ (\bibinfo {year} {2016})\ p.\ \bibinfo {pages} {TUYA01}\BibitemShut {NoStop}%
\bibitem [{\citenamefont {Ablikim}\ \emph {et~al.}(2020{\natexlab{a}})\citenamefont {Ablikim} \emph {et~al.}}]{WhitePaper}%
  \BibitemOpen
  \bibfield  {author} {\bibinfo {author} {\bibfnamefont {M.}~\bibnamefont {Ablikim}} \emph {et~al.} (\bibinfo {collaboration} {BESIII}),\ }\bibfield  {title} {\bibinfo {title} {{Future physics programme of BESIII}},\ }\href {https://doi.org/10.1088/1674-1137/44/4/040001} {\bibfield  {journal} {\bibinfo  {journal} {Chin. Phys. C}\ }\textbf {\bibinfo {volume} {44}},\ \bibinfo {pages} {040001} (\bibinfo {year} {2020}{\natexlab{a}})},\ \Eprint {https://arxiv.org/abs/1912.05983} {arXiv:1912.05983 [hep-ex]} \BibitemShut {NoStop}%
\bibitem [{\citenamefont {Li}\ \emph {et~al.}(2017)\citenamefont {Li} \emph {et~al.}}]{tof1}%
  \BibitemOpen
  \bibfield  {author} {\bibinfo {author} {\bibfnamefont {X.}~\bibnamefont {Li}} \emph {et~al.},\ }\bibfield  {title} {\bibinfo {title} {{Study of MRPC technology for BESIII endcap-TOF upgrade}},\ }\href {https://doi.org/10.1007/s41605-017-0014-2} {\bibfield  {journal} {\bibinfo  {journal} {Radiat. Detect. Technol. Methods}\ }\textbf {\bibinfo {volume} {1}},\ \bibinfo {pages} {13} (\bibinfo {year} {2017})}\BibitemShut {NoStop}%
\bibitem [{\citenamefont {Guo}\ \emph {et~al.}(2017)\citenamefont {Guo} \emph {et~al.}}]{tof2}%
  \BibitemOpen
  \bibfield  {author} {\bibinfo {author} {\bibfnamefont {Y.-X.}\ \bibnamefont {Guo}} \emph {et~al.},\ }\bibfield  {title} {\bibinfo {title} {{The study of time calibration for upgraded end cap TOF of BESIII}},\ }\href {https://doi.org/10.1007/s41605-017-0012-4} {\bibfield  {journal} {\bibinfo  {journal} {Radiat. Detect. Technol. Methods}\ }\textbf {\bibinfo {volume} {1}},\ \bibinfo {pages} {15} (\bibinfo {year} {2017})}\BibitemShut {NoStop}%
\bibitem [{\citenamefont {Cao}\ \emph {et~al.}(2020)\citenamefont {Cao} \emph {et~al.}}]{tof3}%
  \BibitemOpen
  \bibfield  {author} {\bibinfo {author} {\bibfnamefont {P.}~\bibnamefont {Cao}} \emph {et~al.},\ }\bibfield  {title} {\bibinfo {title} {{Design and construction of the new BESIII endcap time-of-flight system with MRPC technology}},\ }\href {https://doi.org/10.1016/j.nima.2019.163053} {\bibfield  {journal} {\bibinfo  {journal} {Nucl. Instrum. Meth. A}\ }\textbf {\bibinfo {volume} {953}},\ \bibinfo {pages} {163053} (\bibinfo {year} {2020})}\BibitemShut {NoStop}%
\bibitem [{\citenamefont {Agostinelli}\ \emph {et~al.}(2003)\citenamefont {Agostinelli} \emph {et~al.}}]{GEANT4}%
  \BibitemOpen
  \bibfield  {author} {\bibinfo {author} {\bibfnamefont {S.}~\bibnamefont {Agostinelli}} \emph {et~al.} (\bibinfo {collaboration} {GEANT4}),\ }\bibfield  {title} {\bibinfo {title} {{GEANT4--a simulation toolkit}},\ }\href {https://doi.org/10.1016/S0168-9002(03)01368-8} {\bibfield  {journal} {\bibinfo  {journal} {Nucl. Instrum. Meth. A}\ }\textbf {\bibinfo {volume} {506}},\ \bibinfo {pages} {250} (\bibinfo {year} {2003})}\BibitemShut {NoStop}%
\bibitem [{\citenamefont {Jadach}\ \emph {et~al.}(2000)\citenamefont {Jadach}, \citenamefont {Ward},\ and\ \citenamefont {Was}}]{KKMC}%
  \BibitemOpen
  \bibfield  {author} {\bibinfo {author} {\bibfnamefont {S.}~\bibnamefont {Jadach}}, \bibinfo {author} {\bibfnamefont {B.~F.~L.}\ \bibnamefont {Ward}},\ and\ \bibinfo {author} {\bibfnamefont {Z.}~\bibnamefont {Was}},\ }\bibfield  {title} {\bibinfo {title} {{The precision Monte Carlo event generator K K for two fermion final states in \ee collisions}},\ }\href {https://doi.org/10.1016/S0010-4655(00)00048-5} {\bibfield  {journal} {\bibinfo  {journal} {Comput. Phys. Commun.}\ }\textbf {\bibinfo {volume} {130}},\ \bibinfo {pages} {260} (\bibinfo {year} {2000})},\ \Eprint {https://arxiv.org/abs/hep-ph/9912214} {arXiv:hep-ph/9912214} \BibitemShut {NoStop}%
\bibitem [{\citenamefont {Lange}(2001)}]{EVTGEN}%
  \BibitemOpen
  \bibfield  {author} {\bibinfo {author} {\bibfnamefont {D.~J.}\ \bibnamefont {Lange}},\ }\bibfield  {title} {\bibinfo {title} {{The EvtGen particle decay simulation package}},\ }\href {https://doi.org/10.1016/S0168-9002(01)00089-4} {\bibfield  {journal} {\bibinfo  {journal} {Nucl. Instrum. Meth. A}\ }\textbf {\bibinfo {volume} {462}},\ \bibinfo {pages} {152} (\bibinfo {year} {2001})}\BibitemShut {NoStop}%
\bibitem [{\citenamefont {Navas}\ \emph {et~al.}(2024)\citenamefont {Navas} \emph {et~al.}}]{ParticleDataGroup:2024cfk}%
  \BibitemOpen
  \bibfield  {author} {\bibinfo {author} {\bibfnamefont {S.}~\bibnamefont {Navas}} \emph {et~al.} (\bibinfo {collaboration} {Particle Data Group}),\ }\bibfield  {title} {\bibinfo {title} {{Review of particle physics}},\ }\href {https://doi.org/10.1103/PhysRevD.110.030001} {\bibfield  {journal} {\bibinfo  {journal} {Phys. Rev. D}\ }\textbf {\bibinfo {volume} {110}},\ \bibinfo {pages} {030001} (\bibinfo {year} {2024})}\BibitemShut {NoStop}%
\bibitem [{\citenamefont {Chen}\ \emph {et~al.}(2000)\citenamefont {Chen}, \citenamefont {Huang}, \citenamefont {Qi}, \citenamefont {Zhang},\ and\ \citenamefont {Zhu}}]{lundcharm1}%
  \BibitemOpen
  \bibfield  {author} {\bibinfo {author} {\bibfnamefont {J.~C.}\ \bibnamefont {Chen}}, \bibinfo {author} {\bibfnamefont {G.~S.}\ \bibnamefont {Huang}}, \bibinfo {author} {\bibfnamefont {X.~R.}\ \bibnamefont {Qi}}, \bibinfo {author} {\bibfnamefont {D.~H.}\ \bibnamefont {Zhang}},\ and\ \bibinfo {author} {\bibfnamefont {Y.~S.}\ \bibnamefont {Zhu}},\ }\bibfield  {title} {\bibinfo {title} {{Event generator for $J/\psi$ and $\psi(2S)$ decay}},\ }\href {https://doi.org/10.1103/PhysRevD.62.034003} {\bibfield  {journal} {\bibinfo  {journal} {Phys. Rev. D}\ }\textbf {\bibinfo {volume} {62}},\ \bibinfo {pages} {034003} (\bibinfo {year} {2000})}\BibitemShut {NoStop}%
\bibitem [{\citenamefont {Yang}\ \emph {et~al.}(2014)\citenamefont {Yang}, \citenamefont {Ping},\ and\ \citenamefont {Chen}}]{lundcharm2}%
  \BibitemOpen
  \bibfield  {author} {\bibinfo {author} {\bibfnamefont {R.-L.}\ \bibnamefont {Yang}}, \bibinfo {author} {\bibfnamefont {R.-G.}\ \bibnamefont {Ping}},\ and\ \bibinfo {author} {\bibfnamefont {H.}~\bibnamefont {Chen}},\ }\bibfield  {title} {\bibinfo {title} {{Tuning and validation of the lundcharm model with $J/\psi$ decays}},\ }\href {https://doi.org/10.1088/0256-307X/31/6/061301} {\bibfield  {journal} {\bibinfo  {journal} {Chin. Phys. Lett.}\ }\textbf {\bibinfo {volume} {31}},\ \bibinfo {pages} {061301} (\bibinfo {year} {2014})}\BibitemShut {NoStop}%
\bibitem [{\citenamefont {Barberio}\ \emph {et~al.}(1991)\citenamefont {Barberio}, \citenamefont {van Eijk},\ and\ \citenamefont {Was}}]{PHOTOS}%
  \BibitemOpen
  \bibfield  {author} {\bibinfo {author} {\bibfnamefont {E.}~\bibnamefont {Barberio}}, \bibinfo {author} {\bibfnamefont {B.}~\bibnamefont {van Eijk}},\ and\ \bibinfo {author} {\bibfnamefont {Z.}~\bibnamefont {Was}},\ }\bibfield  {title} {\bibinfo {title} {{PHOTOS: A universal Monte Carlo for QED radiative corrections in decays}},\ }\href {https://doi.org/10.1016/0010-4655(91)90012-A} {\bibfield  {journal} {\bibinfo  {journal} {Comput. Phys. Commun.}\ }\textbf {\bibinfo {volume} {66}},\ \bibinfo {pages} {115} (\bibinfo {year} {1991})}\BibitemShut {NoStop}%
\bibitem [{\citenamefont {Ablikim}\ \emph {et~al.}(2025)\citenamefont {Ablikim} \emph {et~al.}}]{pipipiFPlus}%
  \BibitemOpen
  \bibfield  {author} {\bibinfo {author} {\bibfnamefont {M.}~\bibnamefont {Ablikim}} \emph {et~al.} (\bibinfo {collaboration} {BESIII}),\ }\bibfield  {title} {\bibinfo {title} {{Measurements of the $CP$-even fractions of $D^0\rightarrow{}\ensuremath{\pi}^+\ensuremath{\pi}^-\ensuremath{\pi}^0$ and $D^0\rightarrow{}K^+K^-\ensuremath{\pi}^0$ at BESIII}},\ }\href {https://doi.org/10.1103/PhysRevD.111.012007} {\bibfield  {journal} {\bibinfo  {journal} {Phys. Rev. D}\ }\textbf {\bibinfo {volume} {111}},\ \bibinfo {pages} {012007} (\bibinfo {year} {2025})},\ \Eprint {https://arxiv.org/abs/2409.07197} {arXiv:2409.07197 [hep-ex]} \BibitemShut {NoStop}%
\bibitem [{\citenamefont {Ablikim}\ \emph {et~al.}()\citenamefont {Ablikim} \emph {et~al.}}]{CompanionPaper}%
  \BibitemOpen
  \bibfield  {author} {\bibinfo {author} {\bibfnamefont {M.}~\bibnamefont {Ablikim}} \emph {et~al.} (\bibinfo {collaboration} {BESIII}),\ }\bibfield  {title} {\bibinfo {title} {{First observation of quantum correlations in $e^+e^-\to XD\bar{D}$ and $C$-even constrained $D\bar{D}$ pairs}},\ }\href@noop {} {\ }\Eprint {https://arxiv.org/abs/2506.07906} {arXiv:2506.07906 [hep-ex]} \BibitemShut {NoStop}%
\bibitem [{\citenamefont {Parzen}(1962)}]{KDE}%
  \BibitemOpen
  \bibfield  {author} {\bibinfo {author} {\bibfnamefont {E.}~\bibnamefont {Parzen}},\ }\bibfield  {title} {\bibinfo {title} {On estimation of a probability density function and mode},\ }\href {http://www.jstor.org/stable/2237880} {\bibfield  {journal} {\bibinfo  {journal} {The Annals of Mathematical Statistics}\ }\textbf {\bibinfo {volume} {33}},\ \bibinfo {pages} {1065} (\bibinfo {year} {1962})}\BibitemShut {NoStop}%
\bibitem [{\citenamefont {Ablikim}\ \emph {et~al.}(2020{\natexlab{b}})\citenamefont {Ablikim} \emph {et~al.}}]{psippKspipiPRD}%
  \BibitemOpen
  \bibfield  {author} {\bibinfo {author} {\bibfnamefont {M.}~\bibnamefont {Ablikim}} \emph {et~al.} (\bibinfo {collaboration} {BESIII}),\ }\bibfield  {title} {\bibinfo {title} {{Model-independent determination of the relative strong-phase difference between $D^0$ and $\bar{D}^0\rightarrow K^0_{S,L}\pi^+\pi^-$ and its impact on the measurement of the CKM angle $\gamma/\phi_3$}},\ }\href {https://doi.org/10.1103/PhysRevD.101.112002} {\bibfield  {journal} {\bibinfo  {journal} {Phys. Rev. D}\ }\textbf {\bibinfo {volume} {101}},\ \bibinfo {pages} {112002} (\bibinfo {year} {2020}{\natexlab{b}})},\ \Eprint {https://arxiv.org/abs/2003.00091} {arXiv:2003.00091 [hep-ex]} \BibitemShut {NoStop}%
\bibitem [{\citenamefont {Ablikim}\ \emph {et~al.}(2020{\natexlab{c}})\citenamefont {Ablikim} \emph {et~al.}}]{psippKspipiPRL}%
  \BibitemOpen
  \bibfield  {author} {\bibinfo {author} {\bibfnamefont {M.}~\bibnamefont {Ablikim}} \emph {et~al.} (\bibinfo {collaboration} {BESIII}),\ }\bibfield  {title} {\bibinfo {title} {{Determination of Strong-Phase Parameters in $D\rightarrow K^0_{S,L}\pi^+\pi^-$}},\ }\href {https://doi.org/10.1103/PhysRevLett.124.241802} {\bibfield  {journal} {\bibinfo  {journal} {Phys. Rev. Lett.}\ }\textbf {\bibinfo {volume} {124}},\ \bibinfo {pages} {241802} (\bibinfo {year} {2020}{\natexlab{c}})},\ \Eprint {https://arxiv.org/abs/2002.12791} {arXiv:2002.12791 [hep-ex]} \BibitemShut {NoStop}%
\bibitem [{\citenamefont {Ablikim}\ \emph {et~al.}(2022{\natexlab{b}})\citenamefont {Ablikim} \emph {et~al.}}]{BESIII:2021yvc}%
  \BibitemOpen
  \bibfield  {author} {\bibinfo {author} {\bibfnamefont {M.}~\bibnamefont {Ablikim}} \emph {et~al.} (\bibinfo {collaboration} {BESIII}),\ }\bibfield  {title} {\bibinfo {title} {{Cross section measurements of the $e^+e^-\to D^{*+}D^{*-}$ and $e^+e^-\to D^{*+}D^{-}$ processes at center-of-mass energies from 4.085 to 4.600 GeV}},\ }\href {https://doi.org/10.1007/JHEP05(2022)155} {\bibfield  {journal} {\bibinfo  {journal} {JHEP}\ }\textbf {\bibinfo {volume} {05}},\ \bibinfo {pages} {155}},\ \Eprint {https://arxiv.org/abs/2112.06477} {arXiv:2112.06477 [hep-ex]} \BibitemShut {NoStop}%
\bibitem [{\citenamefont {Skwarnicki}(1986)}]{Skwarnicki:1986xj}%
  \BibitemOpen
  \bibfield  {author} {\bibinfo {author} {\bibfnamefont {T.}~\bibnamefont {Skwarnicki}},\ }\emph {\bibinfo {title} {{A study of the radiative CASCADE transitions between the Upsilon-Prime and Upsilon resonances}}},\ \href@noop {} {Ph.D. thesis},\ \bibinfo  {school} {Cracow, INP} (\bibinfo {year} {1986})\BibitemShut {NoStop}%
\end{thebibliography}%

\end{document}